\documentclass[10pt]{article}
\usepackage{fullpage}
\usepackage{amsmath}
\usepackage{amssymb}
\usepackage{amsfonts}
\usepackage{comment}
\usepackage[dvips]{epsfig}
\usepackage{color}
\usepackage{cite}

\def\red{\textcolor{black}}

\def\##1{\underline{#1}}
\def\=#1{\underline{\underline{#1}}}

\def\+
#1{\underline{\bf #1}}
\def\*#1{\underline{\underline{\bf #1}}}

\def\r#1{(\ref{#1})}
\def\l#1{\label{#1}}
\def\c#1{\cite{#1}}

\def\le{\left(}
\def\ri{\right)}
\def\les{\left[}
\def\ris{\right]}
\def\lec{\left\{}
\def\ric{\right\}}

\def\.{\mbox{ \tiny{$^\bullet$} }}

\def\eps{\varepsilon}

\def\epso{\eps_{\scriptscriptstyle 0}}
\def\lambdao{\lambda_{\scriptscriptstyle 0}}
\def\muo{\mu_{\scriptscriptstyle 0}}

\def\ko{k_{\scriptscriptstyle 0}}

\def\ux{\hat{\#u}_x}
\def\uy{\hat{\#u}_y}
\def\uz{\hat{\#u}_z}

\def\calA{{\cal A}}
\def\calB{{\cal B}}

\def\alphaAone{\alpha_{\calA 1}}
\def\alphaAtwo{\alpha_{\calA 2}}
\def\alphaB{\alpha_{\calB}}

\begin{document}

\begin{center}

\LARGE{ {\bf  Surface waves with negative phase velocity supported by temperature-dependent  hyperbolic  materials
}}
\end{center}
\begin{center}
\vspace{10mm} \large

 Tom G. Mackay\footnote{E--mail: T.Mackay@ed.ac.uk.}\\
{\em School of Mathematics and
   Maxwell Institute for Mathematical Sciences\\
University of Edinburgh, Edinburgh EH9 3FD, UK}\\
and\\
 {\em NanoMM~---~Nanoengineered Metamaterials Group\\ Department of Engineering Science and Mechanics\\
Pennsylvania State University, University Park, PA 16802--6812,
USA}\\
 \vspace{3mm}
 Akhlesh  Lakhtakia\footnote{E--mail: akhlesh@psu.edu}\\
 {\em NanoMM~---~Nanoengineered Metamaterials Group\\ Department of Engineering Science and Mechanics\\
Pennsylvania State University, University Park, PA 16802--6812, USA}

\normalsize

\end{center}

\begin{center}
\vspace{15mm} {\bf Abstract}
\end{center}
A numerical investigation was undertaken to elucidate the propagation of  electromagnetic surface waves guided by the planar interface of two temperature-sensitive materials. One 
partnering 
material
was chosen to be isotropic and the other to be anisotropic. Both partnering  materials were engineered composite materials, based on the temperature-sensitive semiconductor  InSb. At low temperatures the anisotropic partnering material  is a non-hyperbolic uniaxial material; as the temperature is raised this material becomes a hyperbolic uniaxial material. At low temperatures, a solitary  Dyakonov wave propagates along any specific direction in a range of directions parallel to the planar interface. At high temperatures, up to three different surface waves can propagate in certain directions parallel to the planar interface; one of these surface waves propagates with negative phase velocity (NPV).
At a fixed temperature, the range of directions for NPV propagation 
decreases uniformly in extent as the volume fraction of InSb in the isotropic partnering material decreases.  At a fixed volume fraction
of InSb in the isotropic partnering material, the angular range for NPV propagation
varies substantially as the temperature varies.

\vspace{5mm}

\vspace{5mm}

\vspace{10mm}

\section{Introduction}

The planar interface of two dissimilar materials can support the propagation of a  variety of types of surface wave, even when both  partnering materials (on either side of the interface) are    homogeneous, non-magnetic and non-magnetoelectric \c{ESW_book}. For examples, (i) the planar interface of a plasmonic material and a dielectric material can guide the propagation of surface-plasmon-polariton (SPP) waves \c{Pitarke}; and (ii) the planar interface of an isotropic dielectric material and an anisotropic dielectric material can guide  the propagation of Dyakonov waves  { \c{MSS,Dyakonov88,Takayama_exp,DSWreview}}. Generally, SPP waves propagate for wide ranges of directions parallel to the interface plane, for wide ranges of 
constitutive-parameter values of the partnering materials. In contrast, in the case of nondissipative partnering materials, 
  Dyakonov waves are restricted to much smaller ranges of  propagation directions parallel to the interface plane, and only certain restrictive
   ranges of constitutive-parameter values of the partnering materials are necessary for their propagation \red{\c{Walker98,DSWreview,Torner_NN}}. In cases where the partnering materials are dissipative, even to a small extent, Dyakonov-wave propagation is possible for much wider ranges of propagation directions, for much wider ranges  of constitutive-parameter values, and in such cases more than one Dyakonov wave can propagate in a given direction \c{ML_PJ}.

If the constitutive parameters of one (or both) of the partnering materials are  sensitive to \red{temperature,}  then the interface may guide surface waves of different types at different temperatures. In a recent numerical study \c{ML_PJ}, the transition from Dyakonov waves to SPP waves was mediated by varying the temperature for the interface of a temperature-sensitive  isotropic partnering material and a temperature-insensitive anisotropic partnering material. 
The temperature-sensitive material chosen was the semiconductor  InSb. Indeed, InSb may also be used to thermally tune   metamaterials and metasurfaces operating in the terahertz spectral regime \c{InSb_JOpt,Han}.
This ability to control the transition from Dyakonov waves to SPP waves could potentially be harnessed for temperature sensing and thermal imaging applications \c{Rubin,Ekin}.

Hyperbolic materials \c{Smolyaninov}, as exemplified by an anisotropic dielectric material whose permittivity dyadic has a real part with eigenvalues of opposite signs, 
are associated with  exotic phenomenons
such as  negative refraction  \c{Smith04,Schilling,Guan} and the closely related phenomenon of negative phase velocity (NPV)    \c{Depine}. Such materials
 may be exploited in
subwavelength imaging \c{Liu_Science_2007,Li_JAP,Lemoult}, for
 radiative thermal energy transfer \c{Hyperbolic_Planck,Hyperbolic_Guo_OE}, as analogs of curved spacetime \c{Smoly,ML_PRB}, and for
 diffraction gratings capable of directing light into a large number of refraction channels \c{Depine_NJP}, for examples.
 Also, surface-wave propagation supported by hyperbolic materials has  attracted the recent  attention of both theorists and experimentalists \c{Cojocaru,Spoof_SPP,Zhu_JOpt,Peragut,Narimanov_APL,High,Takayama_ACS,Li,Ma,LF1,LF2}.

 Parenthetically, the taxonomy of 
surface waves supported by hyperbolic materials is problematic: 
on the one hand such surface waves could be regarded as Dyakonov waves because of the anisotropy of the hyperbolic material,
 but on the other hand such surface waves could also be regarded as SPP waves because of the negative-valued eigenvalue(s) of the real  part of the hyperbolic material's permittivity dyadic \c{ESW_book}.

In the following sections, we 
numerically
investigate  the propagation of surface waves guided by the interface of two temperature-sensitive  partnering materials. Both
partnering materials are  non-magnetic, \red{non-magnetoelectric,} and engineered materials, one being isotropic and the other  anisotropic.
The anisotropic partnering material is not of the hyperbolic
kind at low temperatures, but it becomes hyperbolic as the temperature rises. Our main result in this paper is
that the high-temperature regime can support surface-wave propagation with NPV, whereas the low-temperature regime cannot. 

In the  notation adopted,
vectors are underlined once and dyadics
twice. 
An $\exp(-i\omega{t})$ dependence on time $t$ is implicit, with $i=\sqrt{-1}$ and angular frequency $\omega$. 
The triad $\lec\ux,\uy, \uz\ric$ contains the  Cartesian
unit vectors, while the position vector is  denoted by $\#r=x\ux+y\uy+z\uz$.
The   free-space wavenumber is  $\ko=\omega\sqrt{\epso\muo}$,  with
 $\epso$ and $\muo$ being the
  permittivity and permeability of free space, respectively;
  and $\lambdao = 2 \pi / \ko$ is the free-space wavelength.

\section{Canonical boundary-value problem} \label{Cbvp}

The setting for our study is the canonical boundary-value problem for surface-wave propagation guided by the planar  interface of a uniaxial  material
and an isotropic  material \c{ESW_book}, with both materials being  non-magnetic and non-magnetoelectric \cite{ODell,ML_PiO}.
As the theory underpinning this  setting is amply described in earlier works  \c{Polo_JNP,ML_PJ},  only the bare essentials need be presented here.

\begin{figure}[!htb]
\begin{center}
\includegraphics[width=12.5cm]{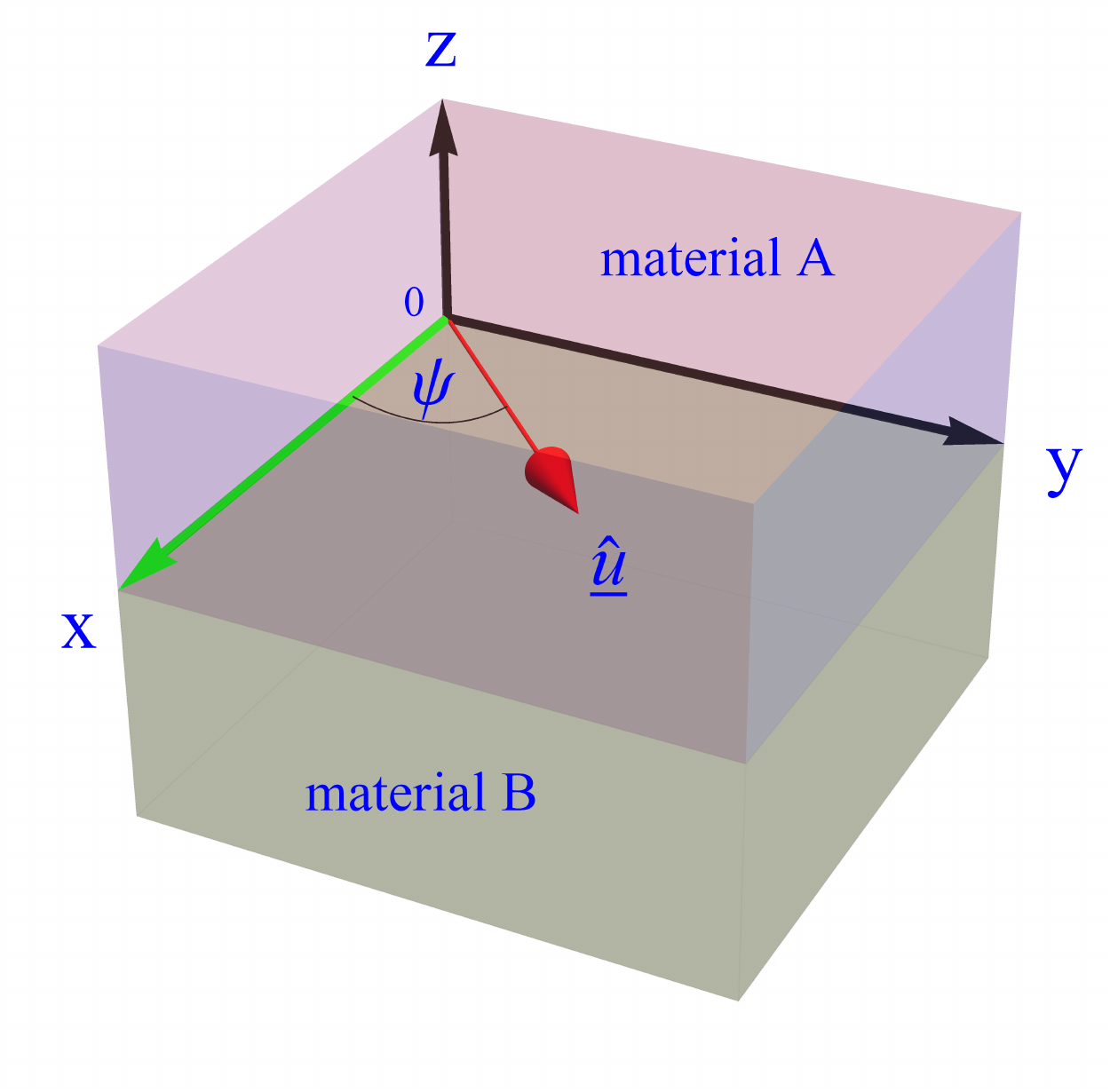}
\end{center}
 \caption{ Schematic representation of the canonical boundary-value problem. Surface-wave propagation is parallel to the $x$ axis, while the optic axis of material $\calA$ is parallel to $\hat{\#{u}}_{\rm }$ which lies in the $xy$ \red{plane at} an angle $\psi$ relative to the $x$ axis.
     } \label{schematic_fig}
\end{figure}

The half-space $z>0$ is occupied by a uniaxial material  labeled $\mathcal{A}$. Its
 relative permittivity
dyadic is given by
\begin{equation}
\=\eps_\mathcal{A} = \eps_\mathcal{A}^{\rm s} \=I + \le
\eps_\mathcal{A}^{\rm t} - \eps_\mathcal{A}^{\rm s} \ri \,
\hat{\#{u}}_{\rm } \, \hat{\#{u}}_{\rm }\,, \l{Ch4_eps_uniaxial}
\end{equation}
wherein  $\=I=\ux\ux+\uy\uy+\uz\uz$ is the identity dyadic \cite{Chen}.
The
optic axis  of material $\calA$, whose direction is
given by the unit vector
\begin{equation}
\hat{\#{u}}_{\rm } = \cos \psi \, \ux +  \sin \psi \, \uy,
\end{equation}
lies in the $xy$ plane,  at an angle $\psi$
relative to the $x$ axis.
The half-space $z<0$ is occupied by  material $\mathcal{B}$ whose relative permittivity
dyadic is $\eps_\mathcal{B}\=I$.  A schematic representation of the canonical boundary-value problem is provided in Fig.~\ref{schematic_fig}.

With no loss of generality, surface-wave propagation parallel to $\ux$ is considered.
The electric field phasor in the half-space $z>0$ is given in terms of  amplitude vectors $\#{\mathcal E}_{\,\calA1}$  and $\#{\mathcal E}_{\,\calA2}$ as
\begin{equation} \label{planewaveA}
 \#E_{\,\calA} (\#r)=  \#{\mathcal E}_{\,\calA1} \,\exp\left({i\#k_{\,\calA1} \cdot\#r}\right)  +
  \#{\mathcal E}_{\,\calA2} \,\exp\left({i\#k_{\,\calA2} \cdot\#r}\right),
\end{equation}
with the wavevectors
\begin{equation} \l{wvA}
\#k_{\, \calA \ell} = \ko \le q \, \ux + i \alpha_{\calA \ell} \uz \ri, \qquad \ell \in \lec 1, 2 \ric,
\end{equation}
and
the decay constants
\begin{equation} \l{a_decay_const}
\left.
\begin{array}{l}
\alpha_{\calA 1} = \sqrt{q^2 -\eps_\calA^s} \vspace{8pt}\\
\alpha_{\calA 2} = \displaystyle{
\sqrt{ \eps^t_\calA \les
q^2 \le \frac{ \cos^2 \psi}{\eps^s_\calA} +
\frac{\sin^2 \psi}{\eps^t_\calA} \ri -1 \ris}}
\end{array}
\right\}\,.
\end{equation}
The electric field phasor in the half-space $z<0$ is given in terms of an amplitude vector $\#{\mathcal E}_{\,\calB}$ as
\begin{equation} \label{planewaveB}
 \#E_{\,\calB} (\#r)=  \#{\mathcal E}_{\,\calB} \,\exp\left({i\#k_{\,\calB} \cdot\#r}\right),
\end{equation}
with the wavevector
 \begin{equation}
\#k_{\, \calB} = \ko \le q \, \ux - i \alpha_\calB \uz \ri\,,
\end{equation}
and decay constant
\begin{equation} \l{b_decay_const}
\alpha_\calB = \sqrt{q^2 - \eps_\calB }.
\end{equation}
Crucially, the inequalities
 ${\rm Re}\lec\alpha_{\calA \ell} \ric>0$, \red{$\ell \in \lec 1,2 \ric$}, and  $\mbox{Re} \lec \alpha_\calB \ric > 0$ must be satisfied   for surface-wave propagation.
 
The normalized propagation constant $q$ is delivered by  solving the
corresponding dispersion relation \c{ML_PJ}
\begin{eqnarray} \l{DE}
&&
  \eps^s_\calA  \le \eps_\calB \sqrt{q^2 - \eps^s_\calA } + \eps^s_\calA \sqrt{q^2 - \eps_\calB} \ri
\les  \sqrt{q^2 - \eps_\calB } + \sqrt{q^2 \le \frac{\eps^t_\calA \cos^2 \psi}{\eps^s_\calA} + \sin^2 \psi \ri - \eps^t_\calA}\,\ris  \tan^2 \psi
\nonumber \\
&& = \sqrt{q^2 - \eps^s_\calA }  \le  \sqrt{q^2 - \eps_\calB } + \sqrt{q^2 - \eps^s_\calA} \ri
\Bigg[ \eps_\calB \le q^2 - \eps^s_\calA \ri +  \left. \eps^s_\calA \sqrt{q^2 - \eps_\calB }  \sqrt{q^2 \le \frac{\eps^t_\calA \cos^2 \psi}{\eps^s_\calA} + \sin^2 \psi \ri - \eps^t_\calA}\,\ris . \nonumber \\ &&
\end{eqnarray}
Numerical methods, such as the Newton--Raphson method \c{N-R}, are generally needed for this task.
 The following symmetries  exist:  if $q = q^\star$ satisfies  Eq.~\r{DE}   at the angle $\psi = \psi^\star$ then $q = q^\star$ also satisfies  Eq.~\r{DE}  
 for $\psi = - \psi^\star$ and $\psi = \pi \pm \psi^\star$. Accordingly, in the proceeding presentation of numerical solutions to Eq.~\r{DE} (in Sec.~\ref{Num_sec}), only the range $0  \leq \psi \leq \pi/2$ need be considered.

\section{Temperature-controlled  partnering materials}

In order 
to control the permittivity parameters of the materials occupying $z>0$ and $z<0$ with temperature,
both half-spaces are taken to be filled with homogenized composite materials containing the temperature-sensitive semiconductor InSb.
The
 relative permittivity of InSb in the terahertz regime is provided by the Drude model \c{Howells,Han} as
\begin{equation}
 \eps_{\text{InSb}} =  \eps_\infty - \frac{\omega_p^2}{\omega^2 + i \gamma \omega }.
\end{equation}
Herein
the plasma frequency $\omega_p = \sqrt{N q_e^2 / 0.015\, \epso \, m_e}$ is determined by the electronic charge $q_e =
-1.60 \times 10^{-19}$~C and mass $m_e = 9.11 \times 10^{-31}$~kg, while 
  the high-frequency relative permittivity $\eps_\infty = 15.68$ and the damping constant $\gamma = \pi \times 10^{11}$~rad~s$^{-1}$.
  The  temperature $T$ (in K) dependence of $ \eps_{\text{InSb}}$ arises via the intrinsic carrier density (in m$^{-3}$) \c{Gruber,Zimpel,Halevi}
\begin{equation}
N = 5.76 \times 10^{20} \, T^{3/2} \, \exp \le - \frac{{\sf E_g}}{2 k_B T} \ri\,,
\end{equation}
which depends upon the bandgap
 ${\sf E_g}=0.26$~eV and the Boltzmann constant
$k_B = 8.62 \times 10^{-5}$ eV $\mbox{K}^{-1}$.

The half-space $z>0$ is taken to be structured as a periodic multilayer, comprising alternating electrically thin sheets of InSb and air, with each sheet being parallel to the interface $z=0$. In the long-wavelength regime, the multilayer functions like a uniaxial continuum whose relative permittivity dyadic has the form given in Eq.~\r{Ch4_eps_uniaxial}.
 The relative permittivity parameters of this homogenized composite material occupying the half-space $z>0$ are estimated using the periodic-multilayer approximation \c{SAR, Krowne} as
\begin{equation}
\left.
\begin{array}{l}
\eps_\calA^s = f^\calA_{\text{InSb}}  \eps_{\text{InSb}} + \le 1-  f^\calA_{\text{InSb}} \ri \vspace{6pt} \\ 
\eps_\calA^t = \les \frac{ f^\calA_{\text{InSb}}}{\eps_{\text{InSb}}} + \le 1-  f^\calA_{\text{InSb}} \ri \ris^{-1}
\end{array}
\right\},
\end{equation}
where $f^\calA_{\text{InSb}} \in \le 0,1 \ri$ is the volume fraction of the InSb layers occupying $z>0$. 
Parenthetically, a multilayer structure with alternate  air layers  can be fabricated by etching 
\c{Grossman,Leclerq},  for example. Also,
the engineered uniaxial material characterized by the relative permittivity dyadic in  Eq.~\r{Ch4_eps_uniaxial} could  arise from the 
homogenization of identically oriented, electrically small, spheroidal particles dispersed randomly \c{Depine,TGM_OE}.  Fabrication  using standard
technologies   has also been reported \c{Takayama_ACS,Li}.

The half-space $z<0$ is taken to be  composed of a random distribution of electrically small spheres made of InSb and a temperature-insensitive polymer, namely high-density polyethylene (HDPE), of relative permittivity $ \eps_{\text{HDPE}}$.
 In the 
long-wavelength regime, the material filling $z<0$ may be regarded as an isotropic  continuum whose relative permittivity is \red{estimated   as}
\begin{equation}
\eps_\calB = f^\calB_{\text{InSb}}  \eps_{\text{InSb}} + \le 1-  f^\calB_{\text{InSb}} \ri  \eps_{\text{HDPE}},
\end{equation}
where $f^\calB_{\text{InSb}} \in \le 0,1 \ri$ is the volume fraction
 of   InSb   in the half-space $z<0$.

\begin{figure}[!htb]
\begin{center}
\includegraphics[width=12.5cm]{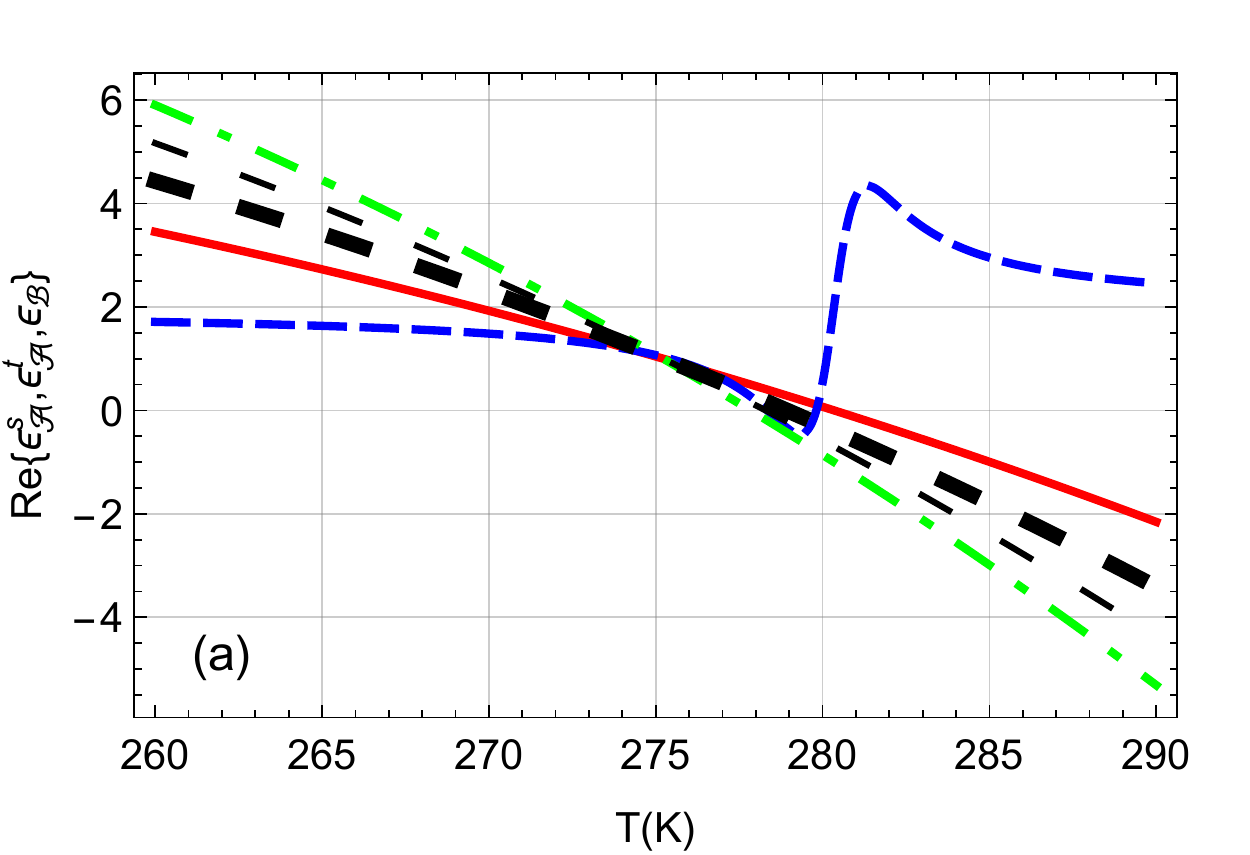}\\
\includegraphics[width=12.5cm]{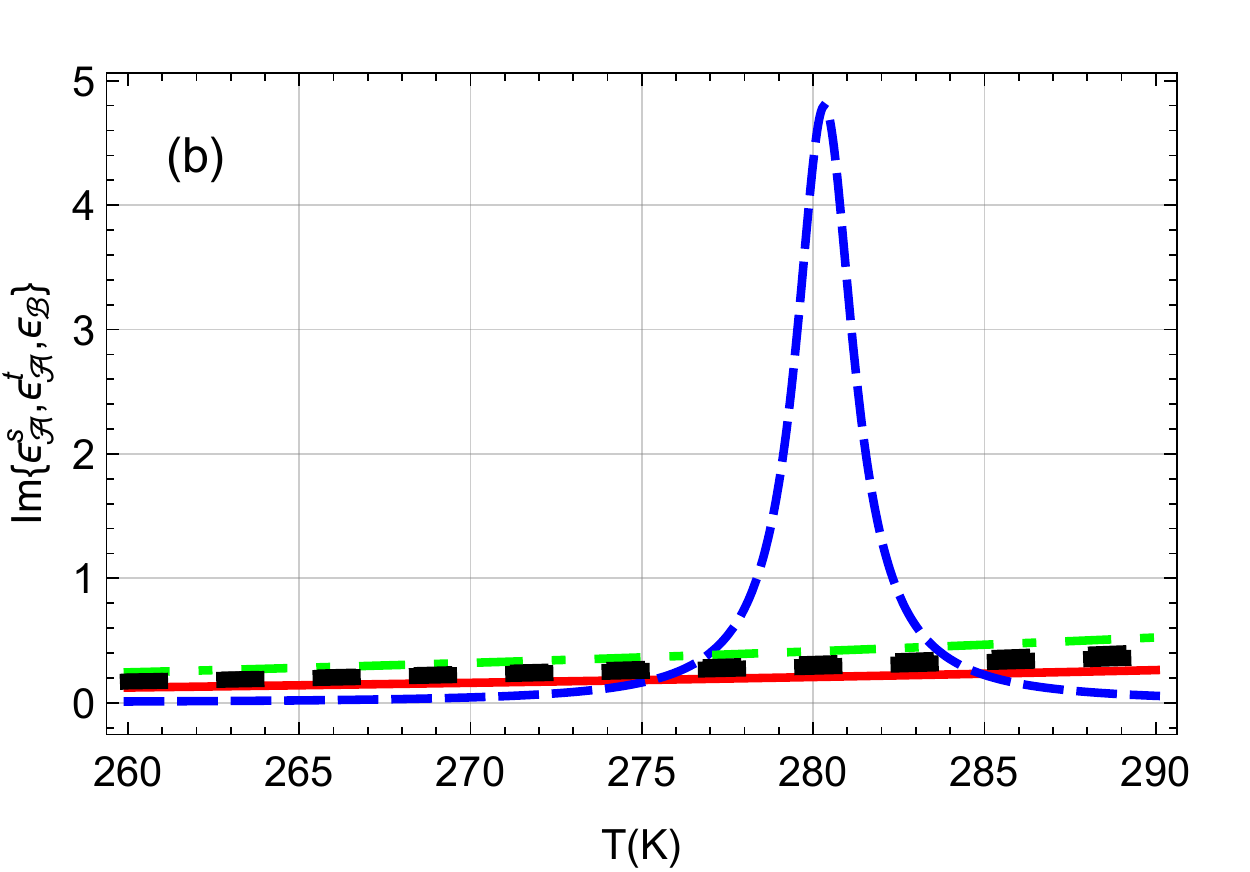}
\end{center}
 \caption{ (a) Real  and (b) imaginary   parts of $\eps^s_\mathcal{A}$ (red solid  curves) and $\eps^t_\mathcal{A}$  (blue dashed curves)  plotted against  $T \in \le 260, 290 \ri $ K, at a frequency of  2.0 THz. Also plotted are  $\eps_\mathcal{B}$ for $f^\calB_{\text{InSb}} = 1$ (green broken dashed  curves),
 $f^\calB_{\text{InSb}} = 0.85$ (thin black dashed  curves, long spaces), and $f^\calB_{\text{InSb}} = 0.7$ (thick black dashed  curves, long spaces). 
     } \label{fig1}
\end{figure}

For the purposes of our   calculations, the volume fraction $f^\calA_{\text{InSb}}  = 0.5$  was selected. The frequency was fixed at $f = 2.0 $ THz (i.e., $\lambdao = 0.15$ mm).
At this frequency, the relative permittivity of HDPE is approximately constant, i.e., $ \eps_{\text{HDPE}}
 \approx  2.387 + 0.006 i $,  over the temperature range $T \in \les 260, 290 \ris$ K  \red{\c{Wietzke}.}
Plots of the real and imaginary parts of the relative permittivity parameters of the partnering materials $\calA$ and $\calB$ 
versus temperature 
are provided in Fig.~\ref{fig1} for the  range $T \in \les 260, 290 \ris$ K with $f^\calB_{\text{InSb}} \in \lec 0.7, 0.85, 1 \ric$.
The real part of    $\eps_\calB$ decreases uniformly as $T$ increases, for the range of values of $f^\calB_{\text{InSb}}$ considered. Specifically, 
for $f^\calB_{\text{InSb}} = 1$, 
$\mbox{Re} \lec \eps_\calB \ric$ is
 positive   for $T <  278$ K,  negative   for $T > 278$ K, and close to
null  at $T= 278$ K. In a similar vein, 
the real part of   $\eps^s_\calA$ decreases uniformly as $T$ increases, with 
$\mbox{Re} \lec \eps^s_\calA \ric$ crossing from positive   to negative   at $T = 280.3$ K.
In contrast, the real part 
 of   $\eps^t_\calA$ is positive   for all values of $T$ considered, except that it is negative in the small
 interval $T \in \le 278.3, 279.8 \ri$ K.
 In particular,  for the temperature range $T \in \le 260, 278.3 \ri \cup \le 279.8, 280.3 \ri $ K the material $\calA$ is not hyperbolic since both
 $\mbox{Re} \lec \eps^s_\calA \ric$ and $\mbox{Re} \lec \eps^t_\calA \ric$ are positive, but for the temperature range
  $T \in \le 278.3, 279.8 \ri \cup \le  280.3, 290 \ri $ K  the material $\calA$ is  hyperbolic since the product 
 $\mbox{Re} \lec \eps^s_\calA \ric \mbox{Re} \lec \eps^t_\calA \ric < 0$ for this  range.
The imaginary parts of $\eps^s_\calA$, $\eps^t_\calA$, and $\eps_\calB$ are generally small, i.e., less than 0.3, across the   temperature range $T \in \les 260, 290 \ris$ K, with the exception of $\mbox{Im} \lec \eps^t_\calA \ric $ which has a localized peak in the vicinity of $T \in \le 276, 284 \ri$ K.

\section{Surface-wave solutions} \l{Num_sec}

In this section representative numerical results are presented of propagation constants, arising as solutions to  Eq.~\r{DE},  and decay constants, as derived from Eqs.~\r{a_decay_const} and \r{b_decay_const}, for surface  waves guided by the interface $z=0$, across the temperature range $T \in \les 260, 290 \ris$ K.

\subsection{Non-hyperbolic material \red{$\calA$}}

\begin{figure}[!htb]
\begin{center}
\includegraphics[width=7.5cm]{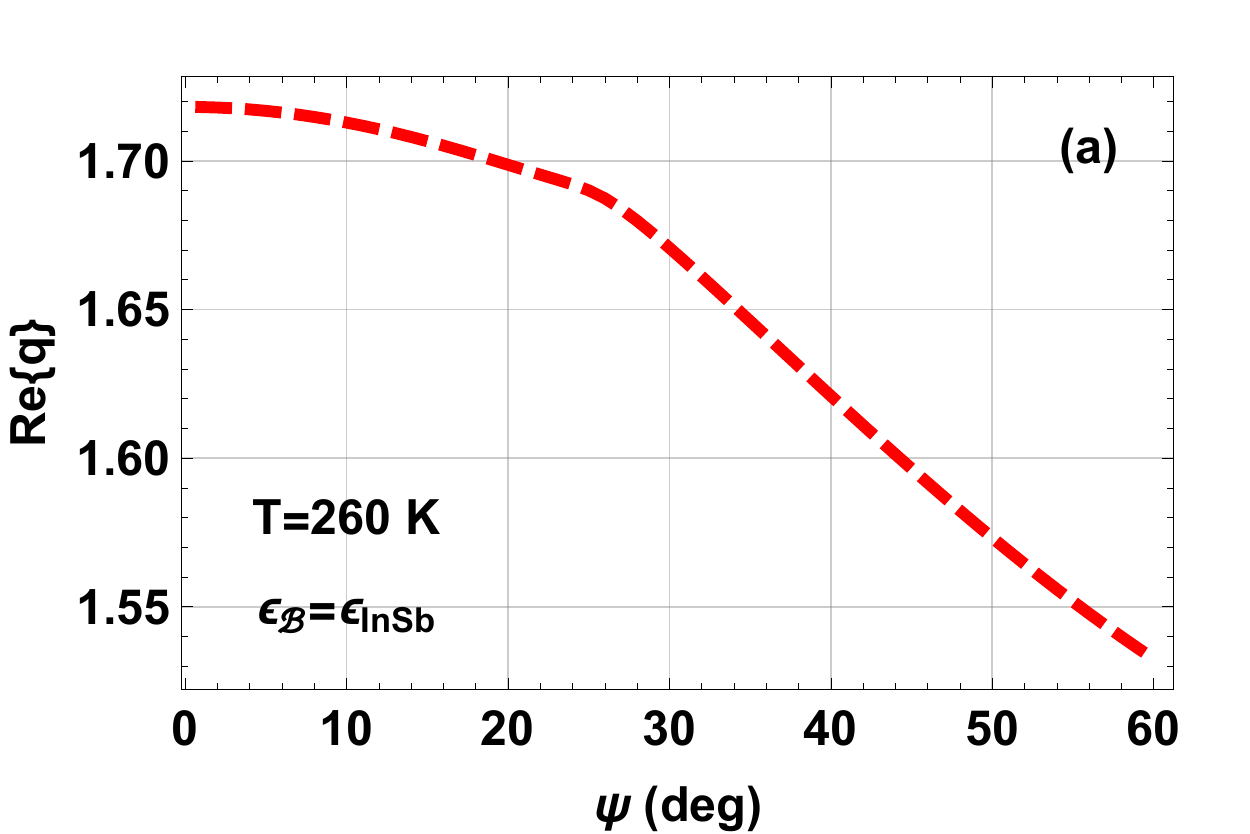} \hfill
\includegraphics[width=7.5cm]{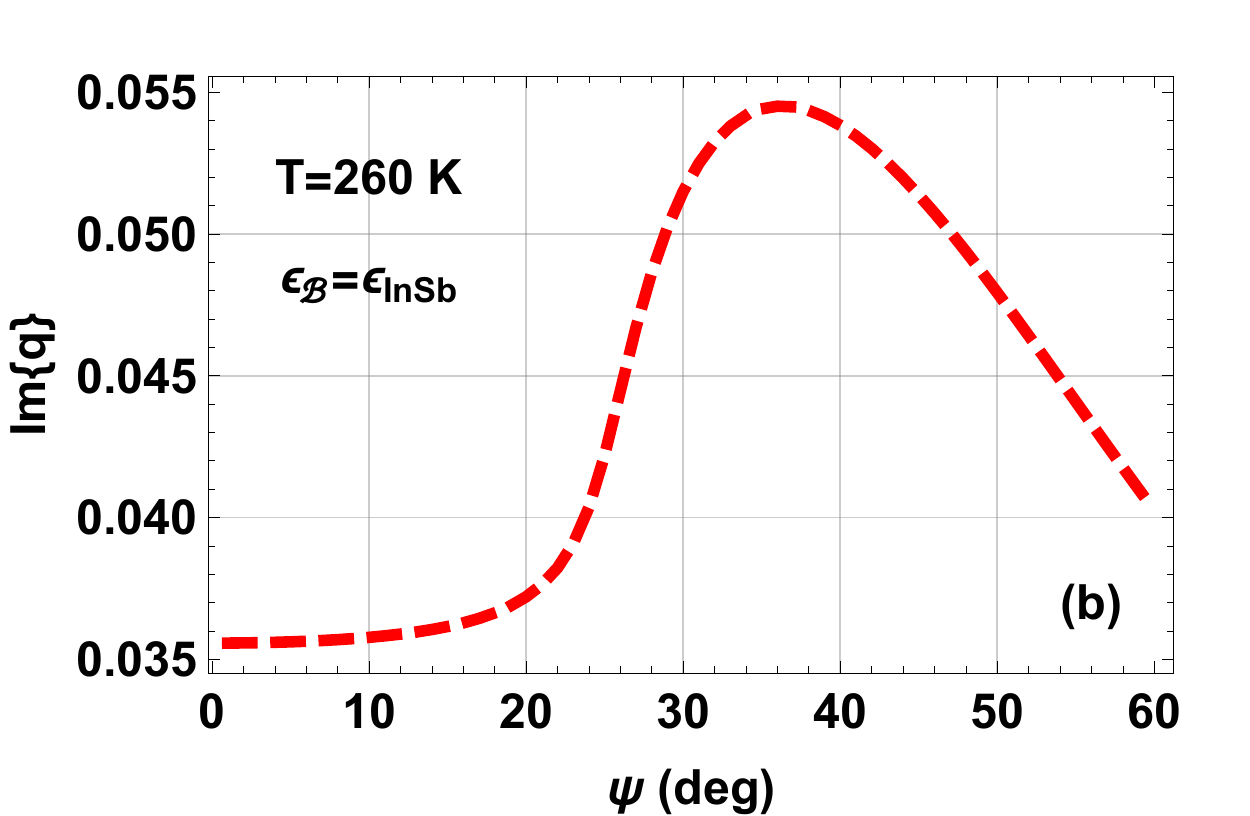} \vspace{4mm} \\
\includegraphics[width=7.5cm]{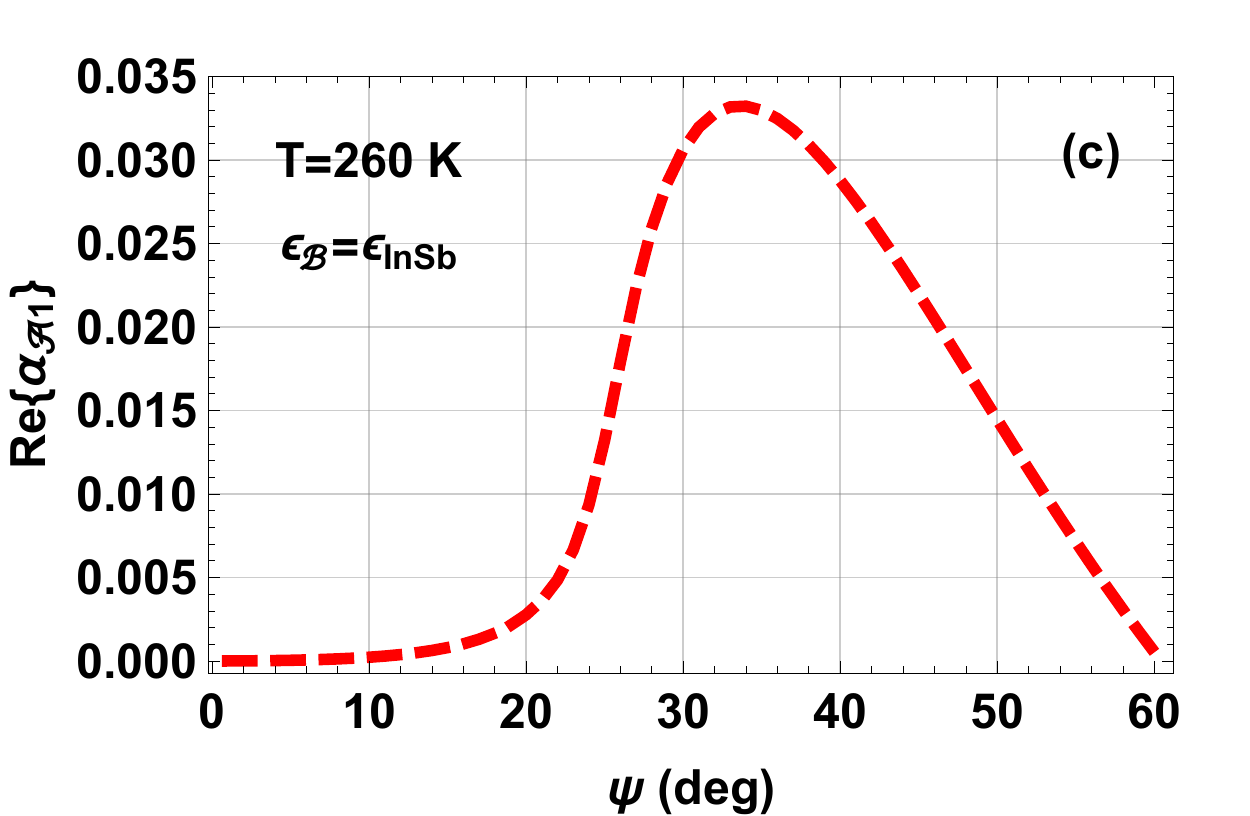} \hfill
\includegraphics[width=7.5cm]{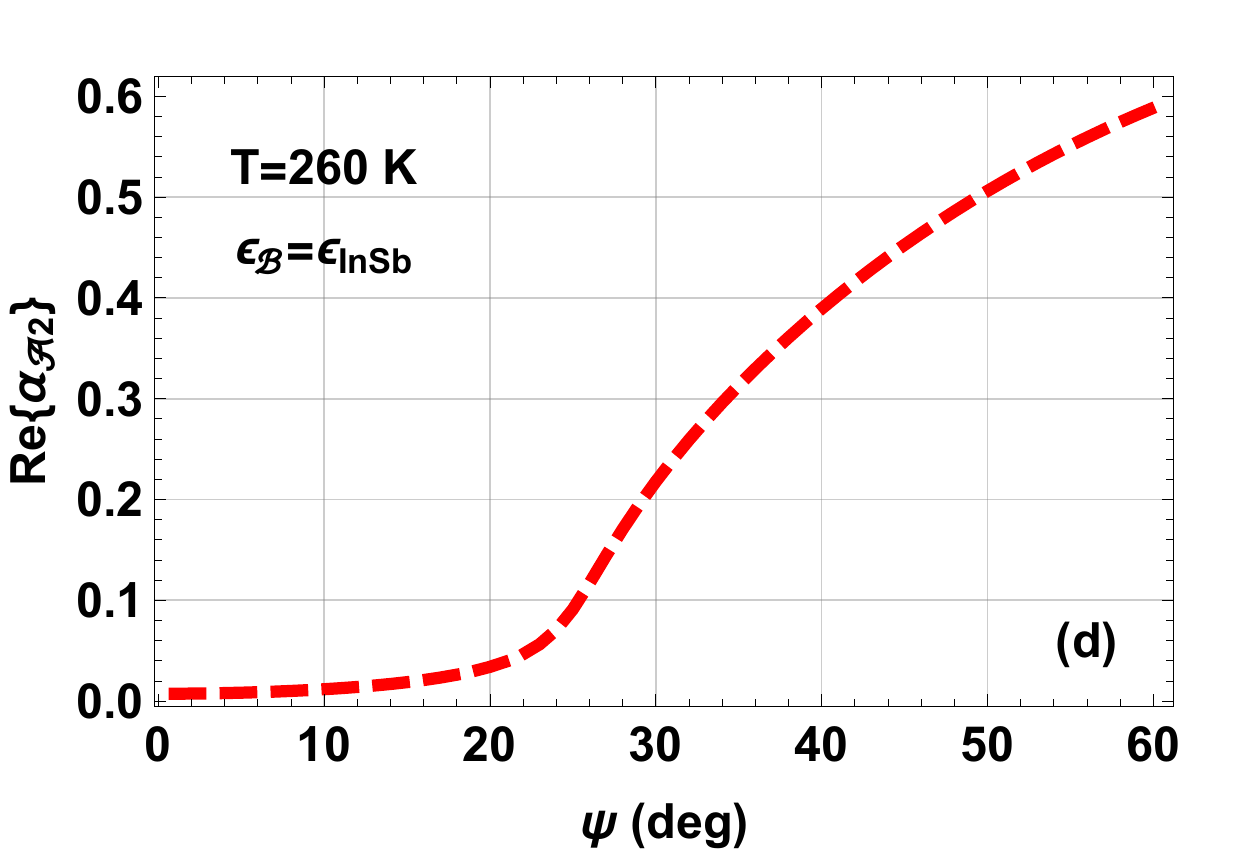} \vspace{4mm} \\
\includegraphics[width=7.5cm]{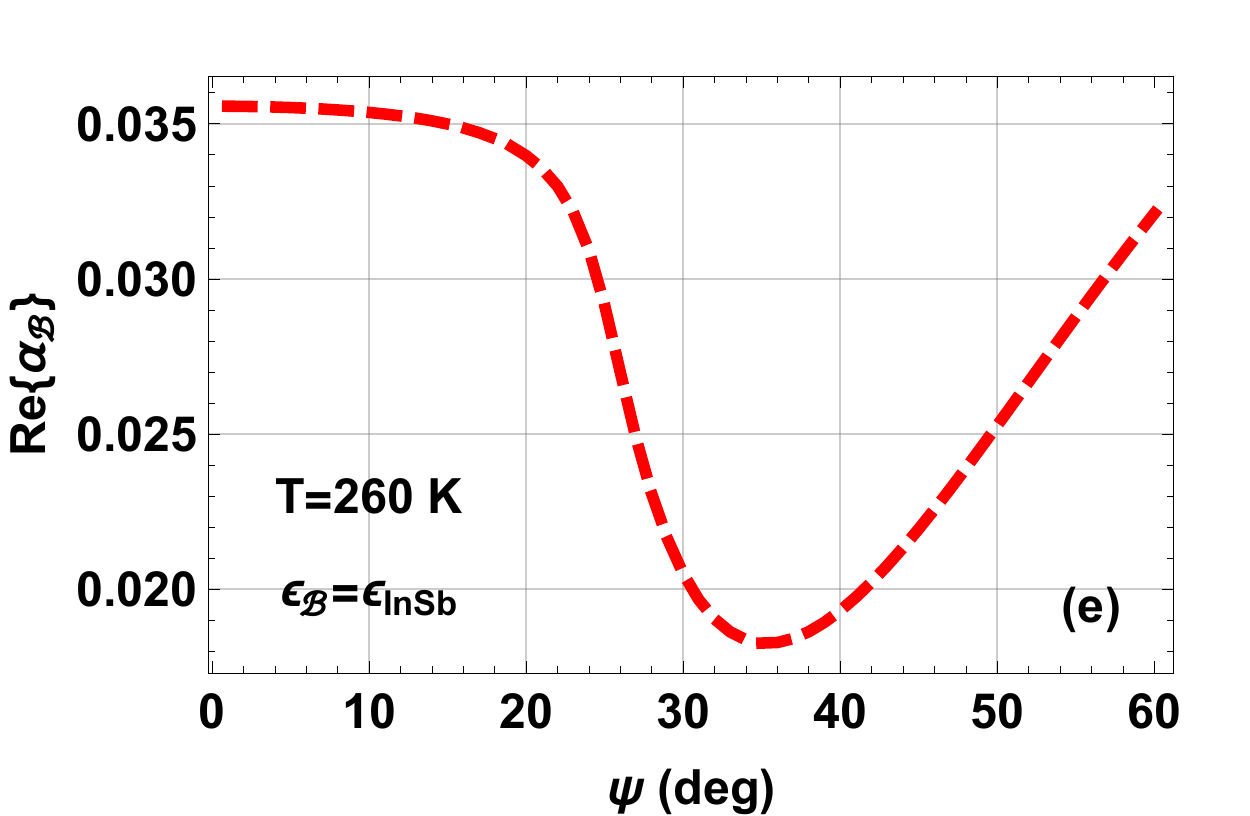} \hfill
\includegraphics[width=7.5cm]{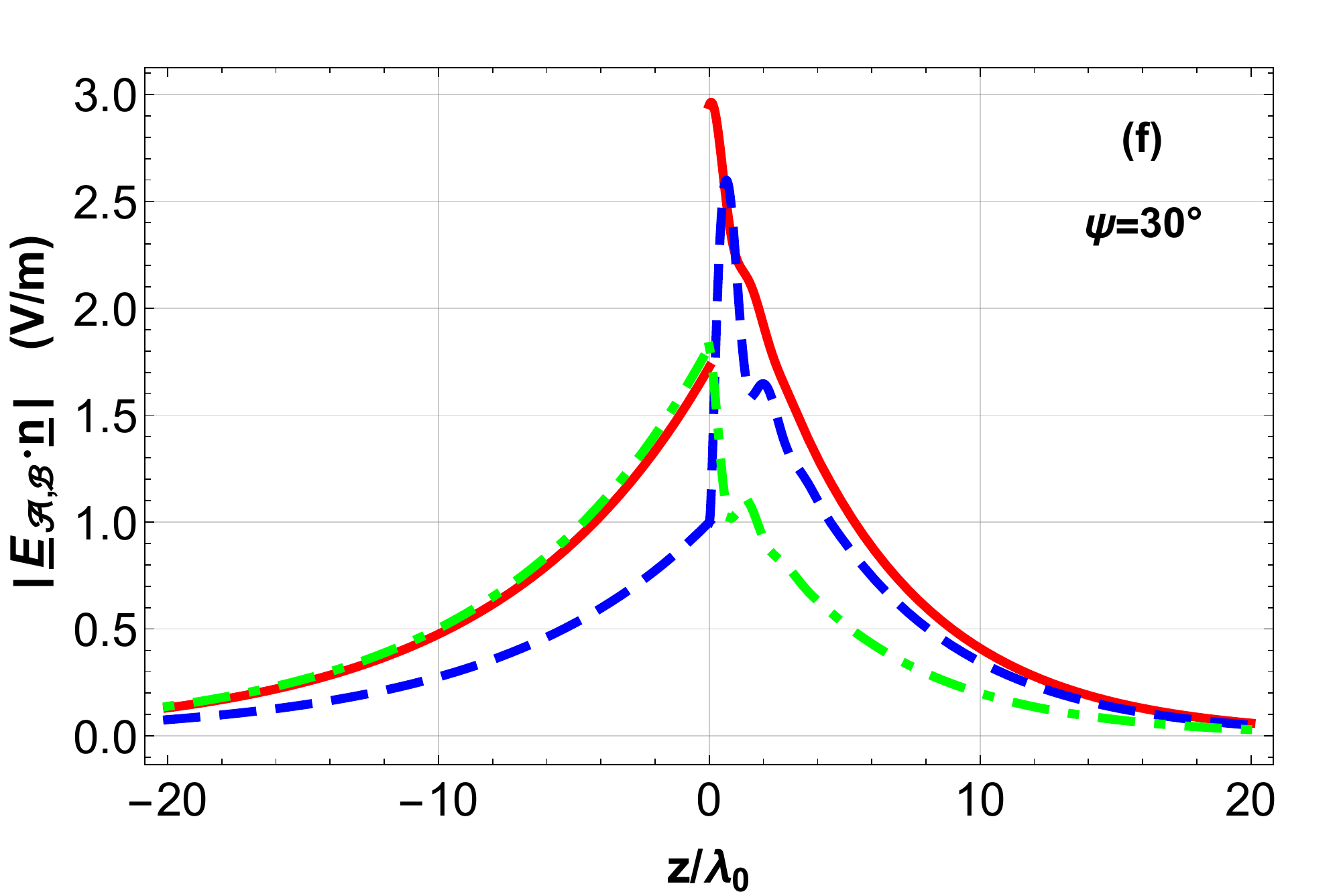}
\end{center}
 \caption{ { (a) ${\rm Re}\lec{q}\ric$, (b) ${\rm Im}\lec{q}\ric$, (c) ${\rm Re}\lec\alpha_{\mathcal{A}1}\ric$,
 (d) ${\rm Re}\lec\alpha_{\mathcal{A}2}\ric$, and (e) ${\rm Re}\lec\alpha_{\mathcal{B}}\ric$} plotted against   $\psi$ for $T= 260$ K with  $f^\calB_{\text{InSb}} = 1$. Here
 $\eps_\mathcal{A}^s =  3.4509 + 0.1222  i $, $\eps_\mathcal{A}^t =  1.7106 + 0.0103 i $, and
  $\eps_\mathcal{B} = 5.9018 + 0.2445   i $. The sole solution   
  exists for $\psi \in \le 0^\circ, 60^\circ \ri$.
  { Also
  the quantities (f) $| \underline{E}_{\, \mathcal{A},\mathcal{B}} (z\hat{\underline{u}}_{\,z}) \. \#n|$ are
 plotted versus $z/\lambdao$, computed for $\psi = 30^\circ$ with $ \#{\mathcal E}_{\,\calB} \. \uy = 1$ V m${}^{-1}$.
 Key for (f):   $\#n = \ux$ broken dashed green curves; $\#n = \uy$ dashed blue curves; $\#n = \uz$ solid red curves.}
  } \label{fig_T260}
\end{figure}

Let us begin our numerical investigation of  surface waves  with the 
temperature regime in which material $\calA$ is not hyperbolic and both materials $\calA$ and $\calB$ are  weakly dissipative dielectric materials with $\mbox{Re} \lec \eps_\calB \ric > \mbox{Re} \lec \eps_\calA^s \ric >  \mbox{Re} \lec \eps_\calA^t \ric > 0 $.
 The real and imaginary parts of $q$, along with the real parts of $\alpha_{\calA1}$, $\alpha_{\calA2}$, and $\alpha_\calB$, are plotted against orientation angle $\psi$ in Fig.~\ref{fig_T260} for $T = 260$ K with  $f^\calB_{\text{InSb}} = 1$.
At this temperature $\eps_\mathcal{A}^s =  3.4509 + 0.1222  i $, $\eps_\mathcal{A}^t =  1.7106 + 0.0103 i $, and
  $\eps_\mathcal{B} = 5.9018 + 0.2445   i $. Over the angular range $0^\circ < \psi < 90^\circ$,  only one solution 
    exists for $\psi \in \le 0^\circ, 60^\circ \ri$. For this solution $\mbox{Re} \lec q \ric > 1 $,  from which it is inferred that the phase speed of this surface wave is smaller than the phase speed of a plane wave in free space. Since both partnering materials are weakly dissipative dielectric materials at this temperature, the  surface wave may be called a Dyakonov wave.
Although the partnering materials are only weakly dissipative, the effects of dissipation are profound. If the corresponding nondissipative case were considered with real-valued permittivity parameters (i.e.,
 $\eps_\mathcal{A}^s =  3.4509  $, $\eps_\mathcal{A}^t =  1.7106 $, and
  $\eps_\mathcal{B} = 5.9018$)
 then no Dyakonov waves would exist, since Dyakonov waves can only be found for nondissipative partnering materials if
   $\eps_\calA^t  >  \eps_\calB  >  \eps_\calA^s$  \cite{Dyakonov88}.
    Furthermore, the angular existence domain of the  solution represented in Fig.~\ref{fig_T260} is much 
    larger than the angular existence domains typically associated with Dyakonov waves guided by the interfaces of nondissipative materials \cite{DSWreview}.
    
 Also provided in Fig.~\ref{fig_T260} are
     profiles of    $|\underline{E}_{\, \ell} (z\uz) \. \#n|$,
$\ell \in \lec \mathcal{A}, \mathcal{B} \ric$,
 versus $z / \lambdao$ with $\#n \in \lec \ux, \uy, \uz \ric$, computed
for  $\psi = 30^\circ$ and $\#{\mathcal E}_{\,\calB} \. \uy = 1$ V m${}^{-1}$. These profiles indicate decay in an approximately exponential manner as the distance 
$\vert{z}\vert$ from the interface $z=0$ increases, in consonance
with  $\mbox{Re}\lec \alphaAone \ric= 0.03062$,  $\mbox{Re} \lec \alphaAtwo \ric =0.21760$, and  $\mbox{Re}\lec \alphaB \ric =0.02050$ all being positive.
This confirms that
these solutions do indeed represent Dyakonov waves that are localized at the interface $z=0$.

\begin{figure}[!htb]
\begin{center}
\includegraphics[width=7.5cm]{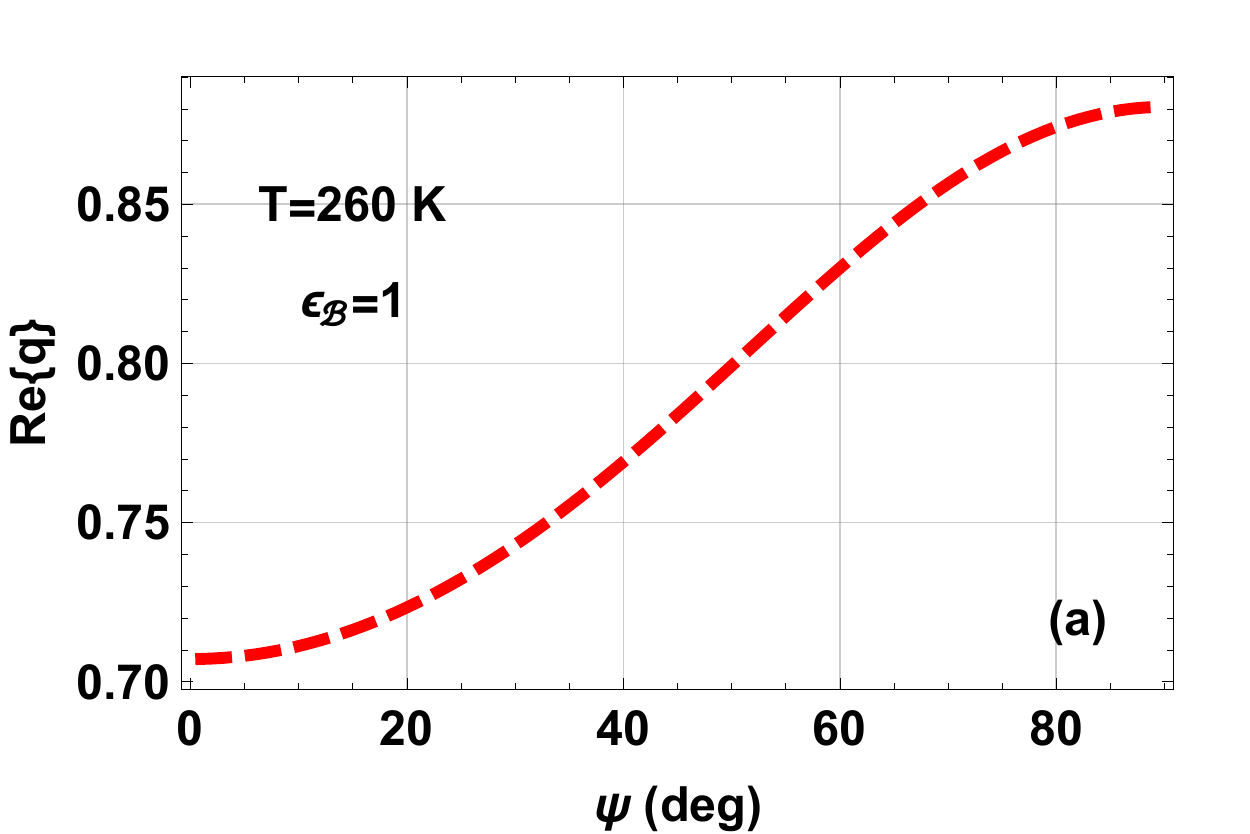} \hfill
\includegraphics[width=7.5cm]{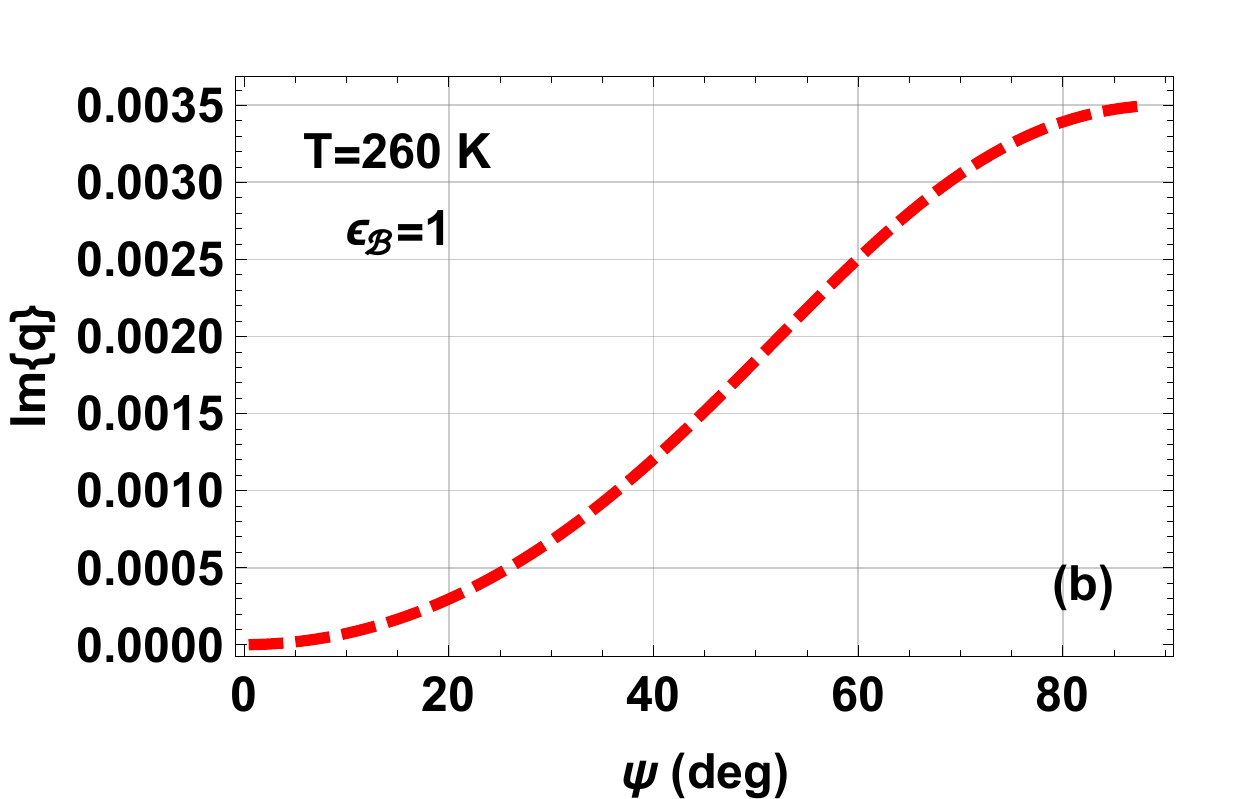} \vspace{4mm} \\
\includegraphics[width=7.5cm]{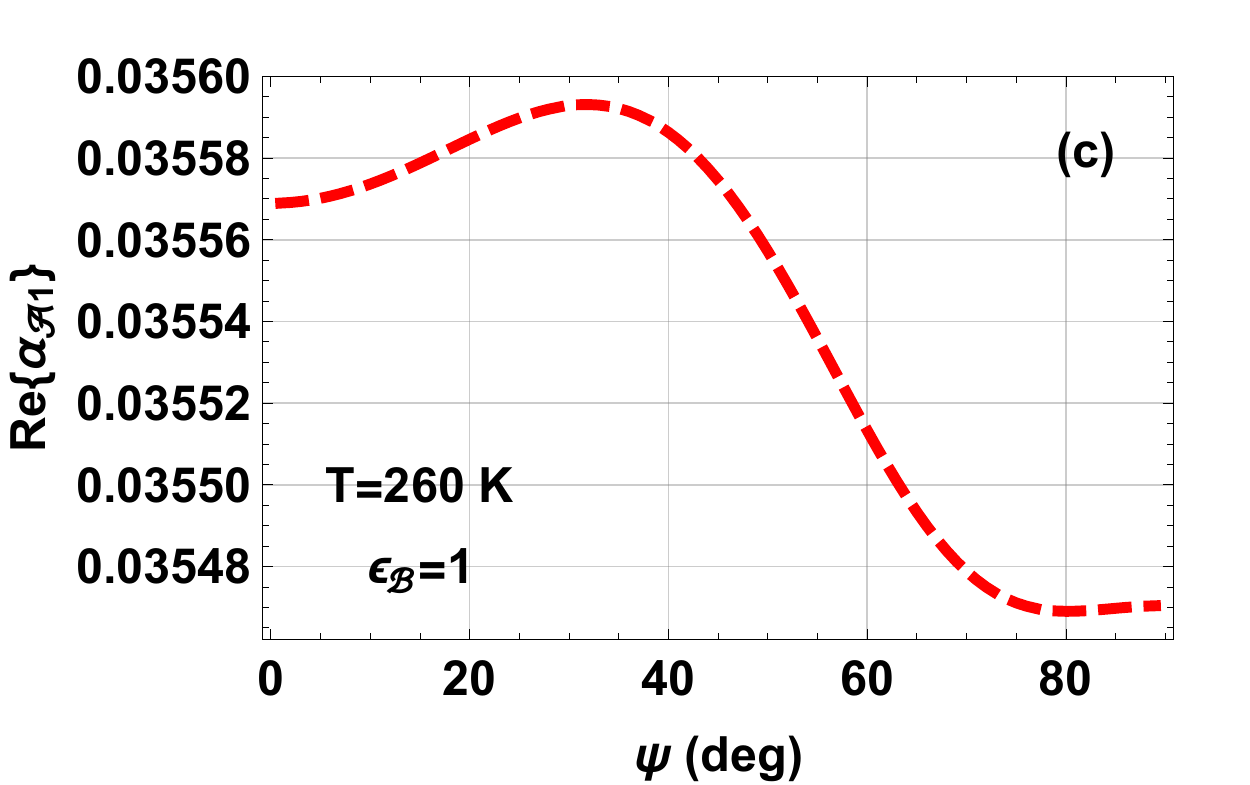} \hfill
\includegraphics[width=7.5cm]{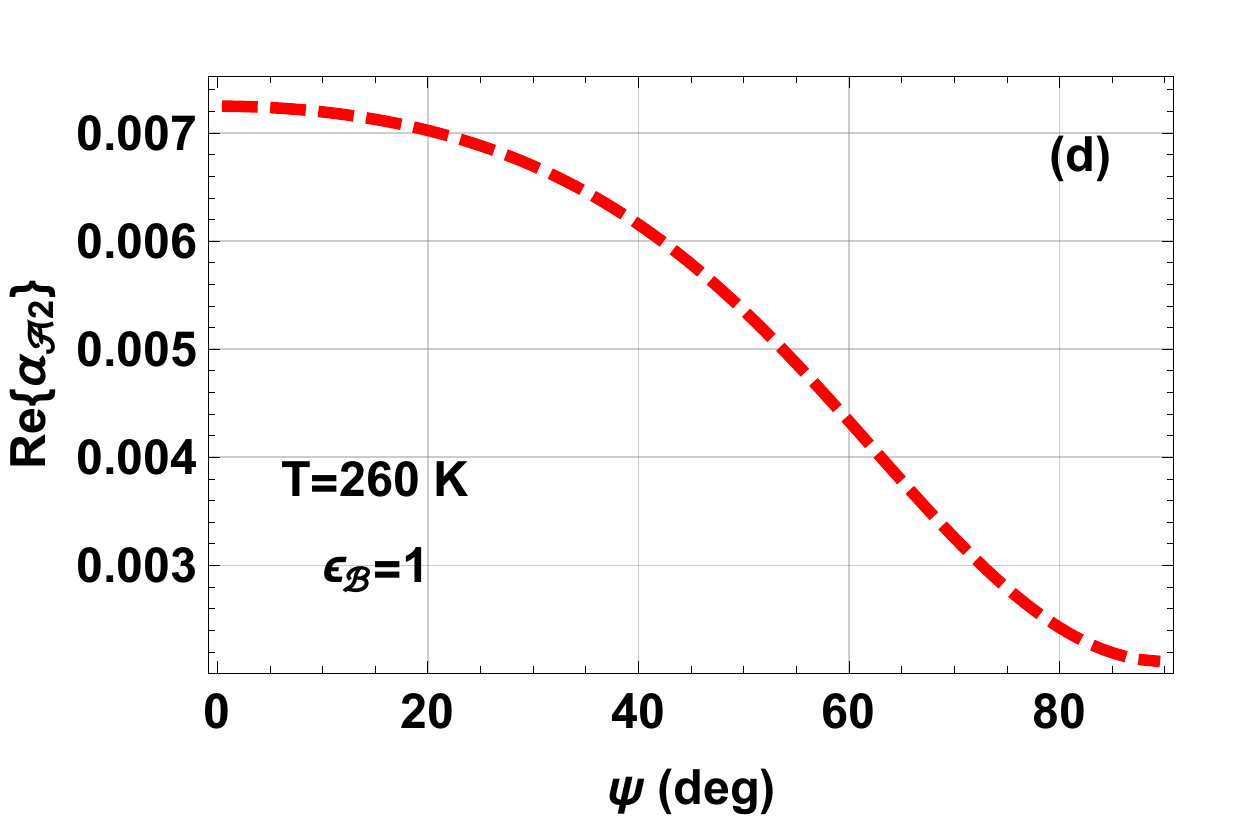} \vspace{4mm} \\
\includegraphics[width=7.5cm]{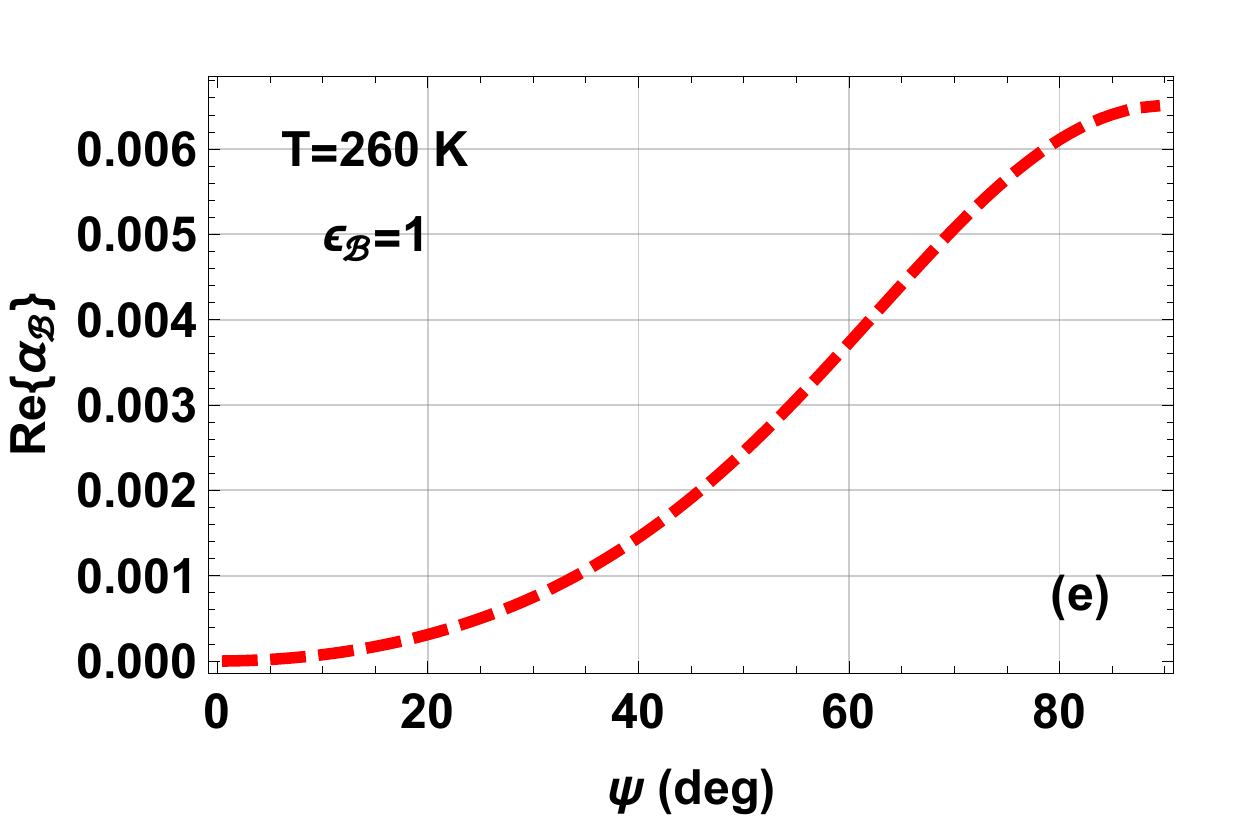} \hfill
\includegraphics[width=7.5cm]{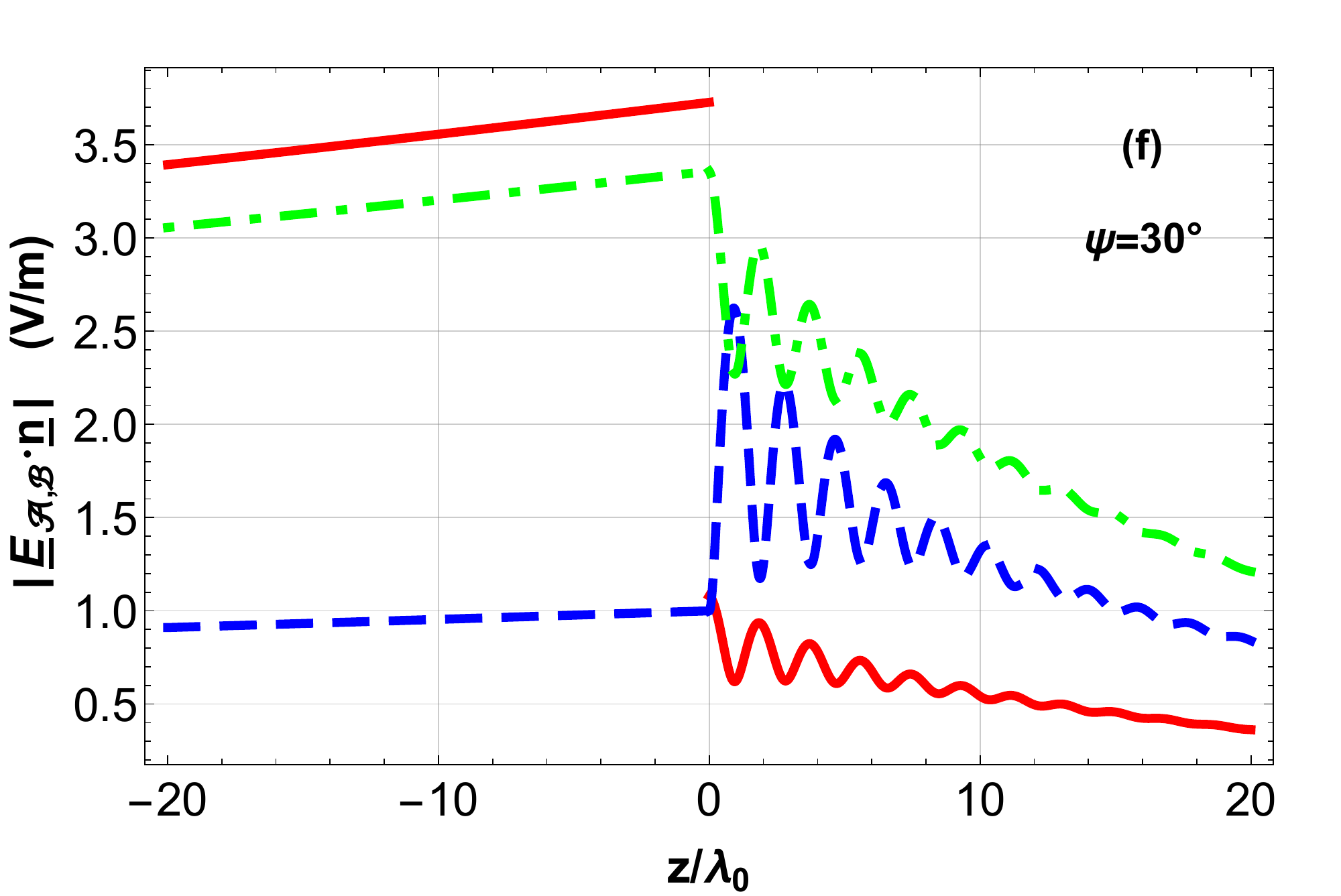}
\end{center}
 \caption{As Fig.~\ref{fig_T260} except that   $\eps_\calB = 1$. The sole solution   
  exists for $\psi \in \le 0^\circ, 90^\circ \ri$.
 } \label{fig_T260_E0}
\end{figure}

By way of comparison and to highlight the effects of InSb in $z<0$, solutions were also computed for the analogous scenario in which the material occupying the half-space $z<0$ is simply air (i.e., $\eps_\calB= 1 $). The corresponding results are presented in Fig.~\ref{fig_T260_E0}. Like the case presented in Fig.~\ref{fig_T260}, there is only one Dyakonov-wave solution. It exists for all  angles $0^\circ < \psi < 90^\circ$.
For this solution $\mbox{Re} \lec q \ric < 1 $ as well.
The imaginary part of the propagation constant $q$ and the real parts of the decay constants $\alpha_{\calA1}$,  $\alpha_{\calA2}$, and  $\alpha_{\calB}$, are all generally smaller for Fig.~\ref{fig_T260_E0} than they are for Fig.~\ref{fig_T260}.

The      profiles of    $|\underline{E}_{\, \ell} (z\uz) \. \#n|$,
$\ell \in \lec \mathcal{A}, \mathcal{B} \ric$,
   provided in Fig.~\ref{fig_T260_E0}(f) 
   reveal that the Dyakonov waves represented in Fig.~\ref{fig_T260_E0}  are less tightly localized to the interface $z=0$  than are
   the Dyakonov waves represented in Fig.~\ref{fig_T260}. Thus, $q=0.74302 + \red{0.00067i}$,
$\mbox{Re} \lec \alphaAone \ric = 0.03559$,  $\mbox{Re} \lec \alphaAtwo \ric = 0.00670$, and  $\mbox{Re} \lec \alphaB \ric = 0.00075$ for  Fig.~\ref{fig_T260_E0}(f),
whereas $q=1.67076 + \red{0.05150i}$, 
 $\mbox{Re}\lec \alphaAone \ric= 0.03062$,  $\mbox{Re} \lec \alphaAtwo \ric =0.21760$, and  $\mbox{Re}\lec \alphaB \ric =0.02050$
 for  Fig.~\ref{fig_T260}(f).

\subsection{Hyperbolic material \red{$\calA$}}

\begin{figure}[!htb]
\begin{center}
\includegraphics[width=7.5cm]{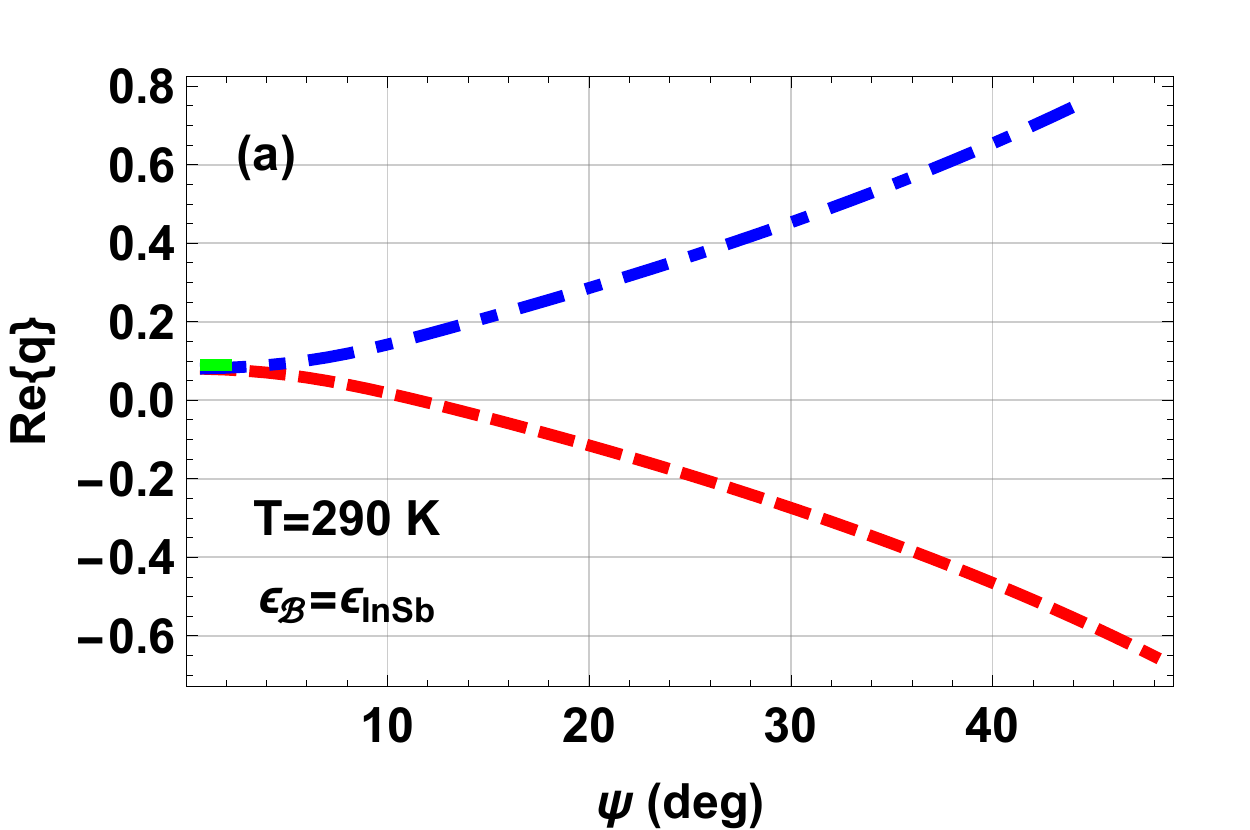} \hfill
\includegraphics[width=7.5cm]{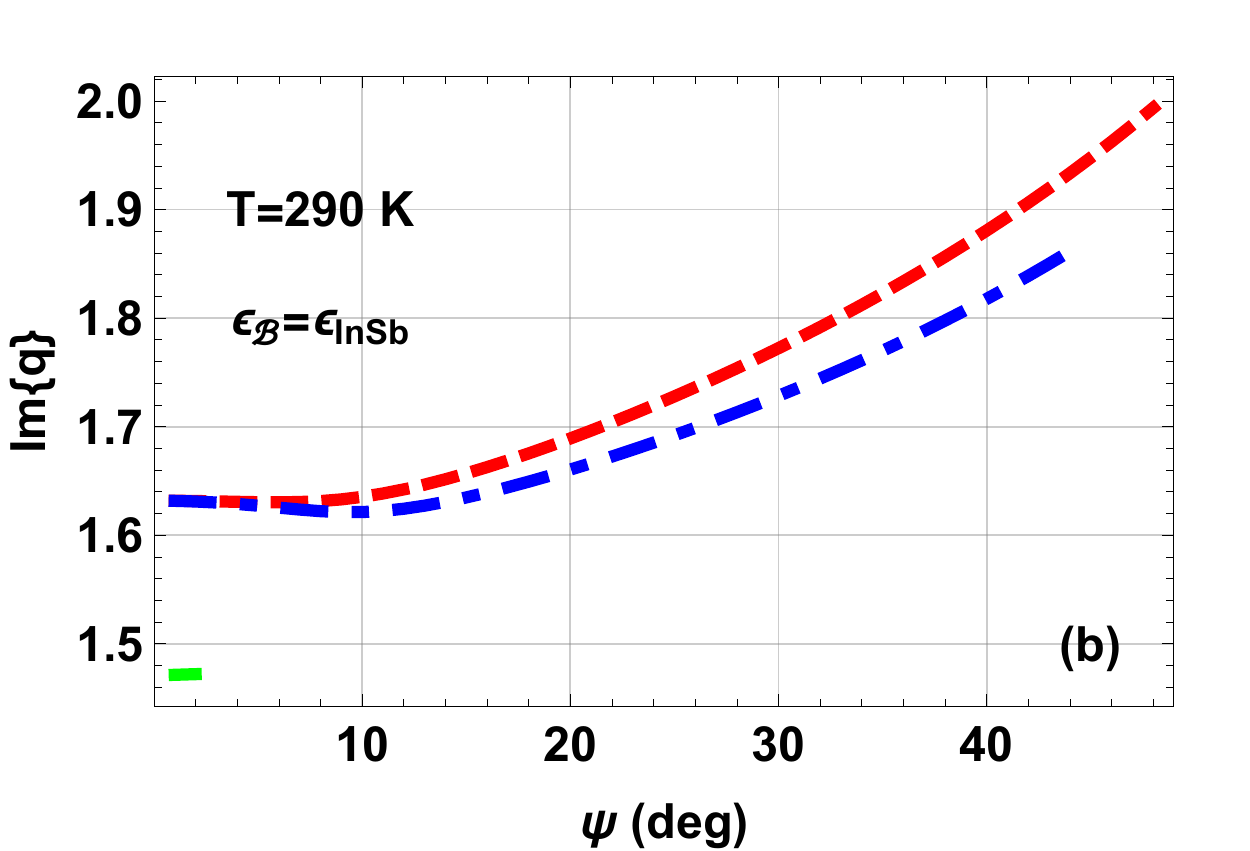} \vspace{4mm} \\
\includegraphics[width=7.5cm]{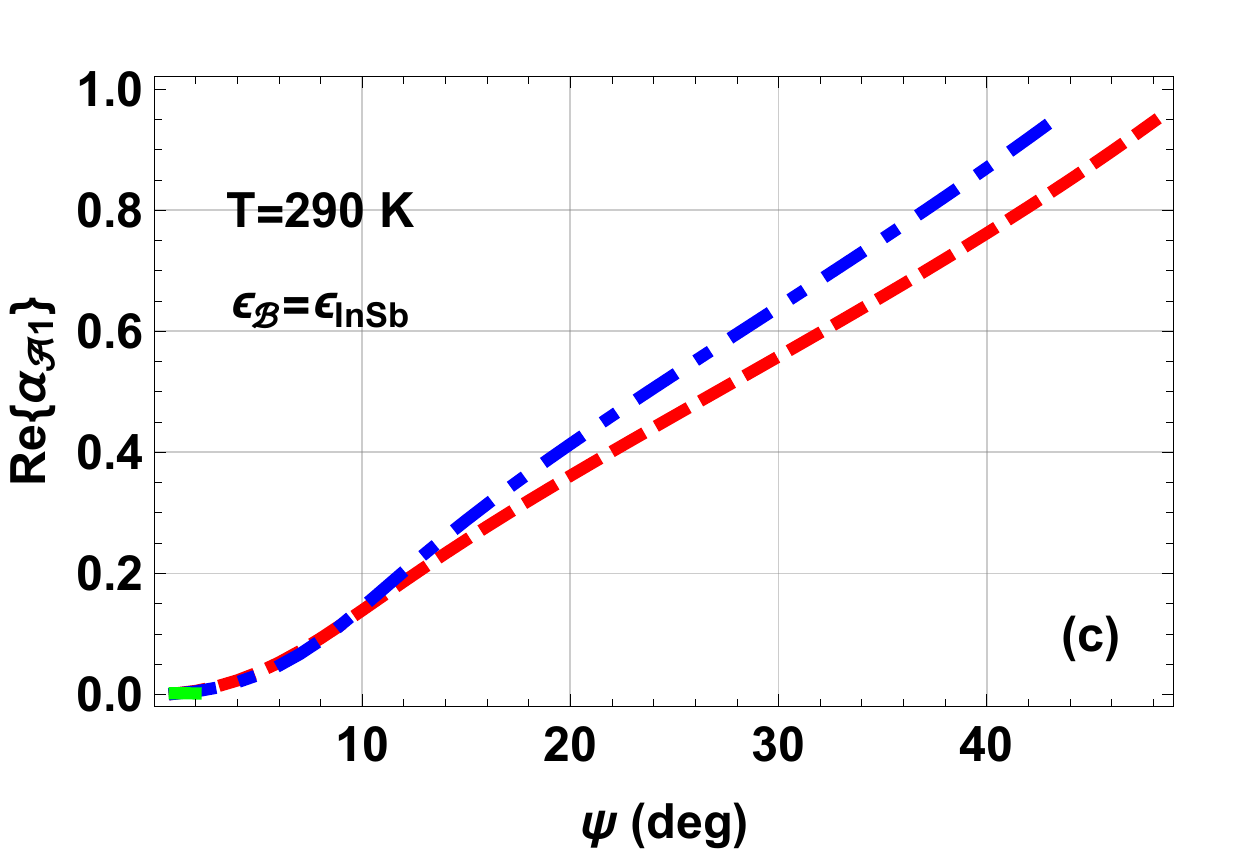} \hfill
\includegraphics[width=7.5cm]{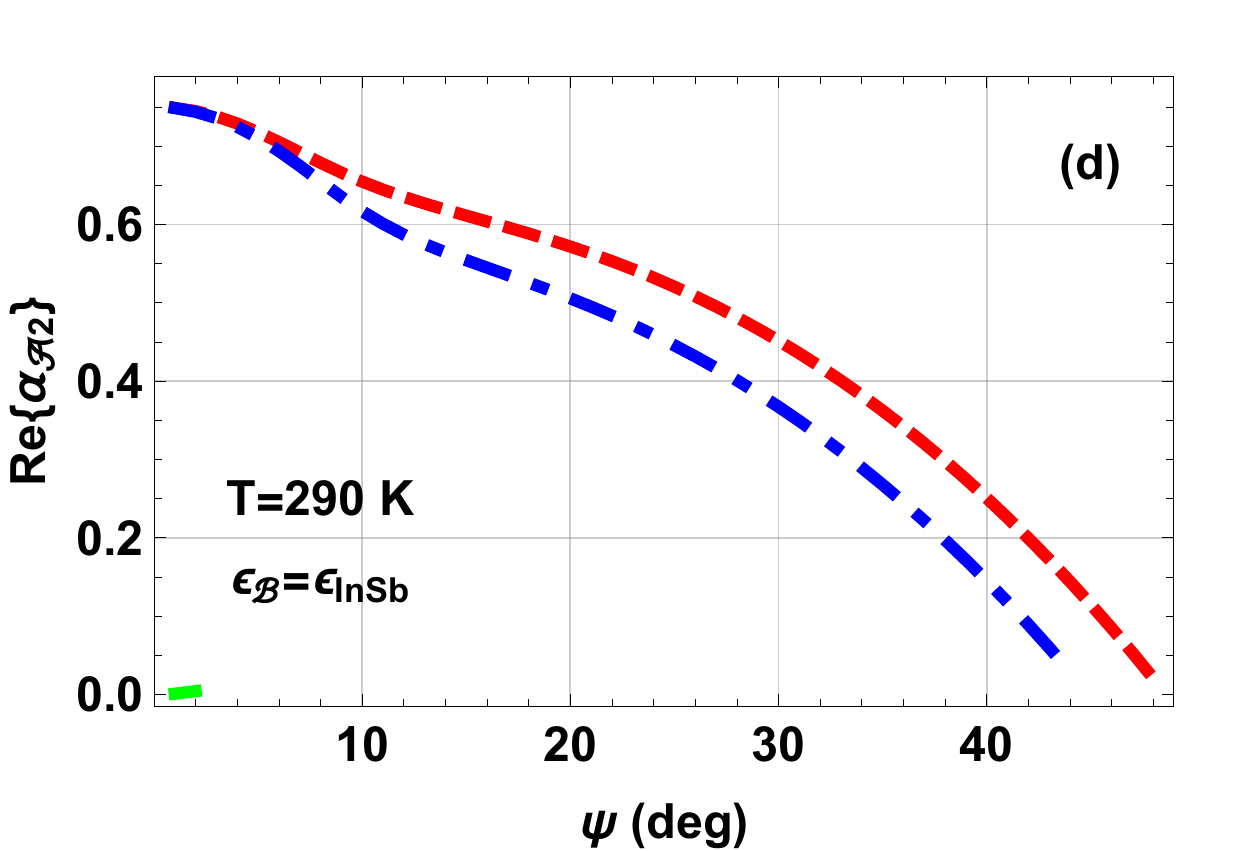} \vspace{4mm} \\
\includegraphics[width=7.5cm]{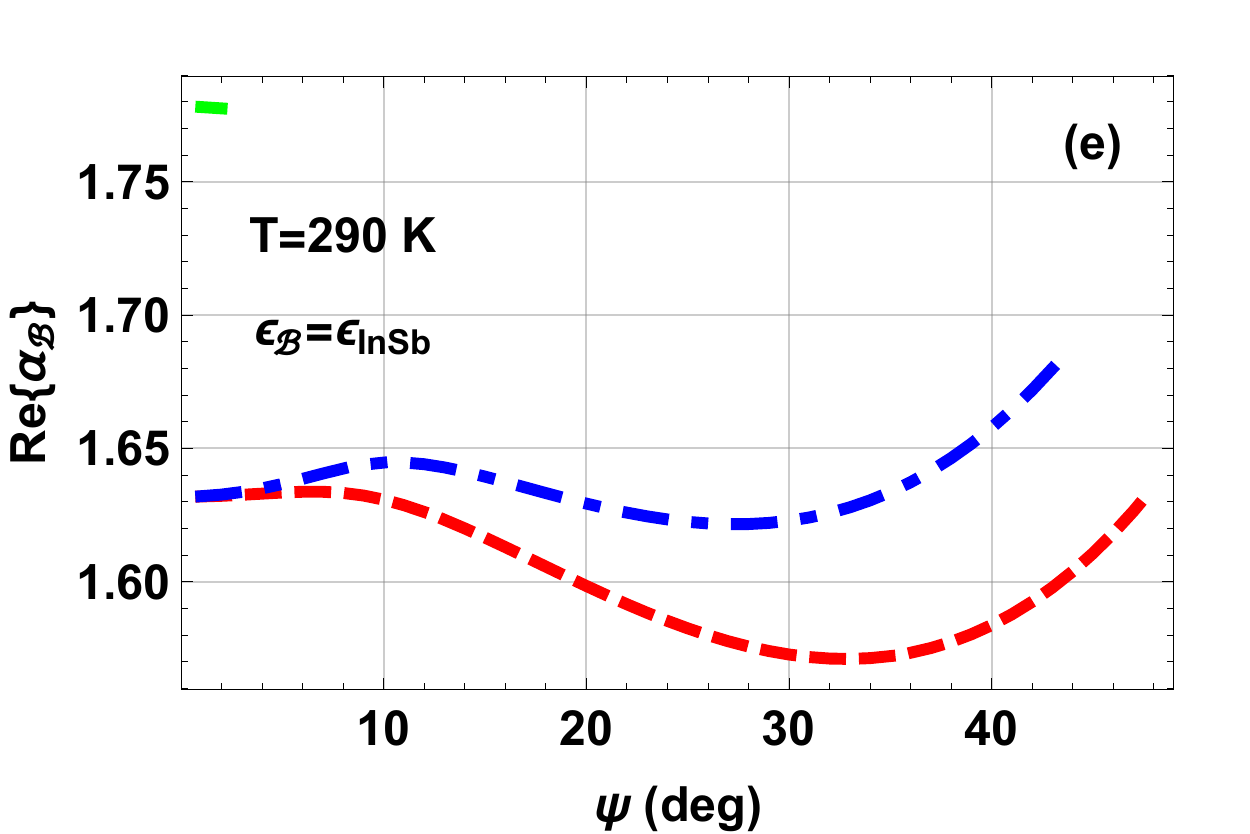}  \hfill
\includegraphics[width=7.5cm]{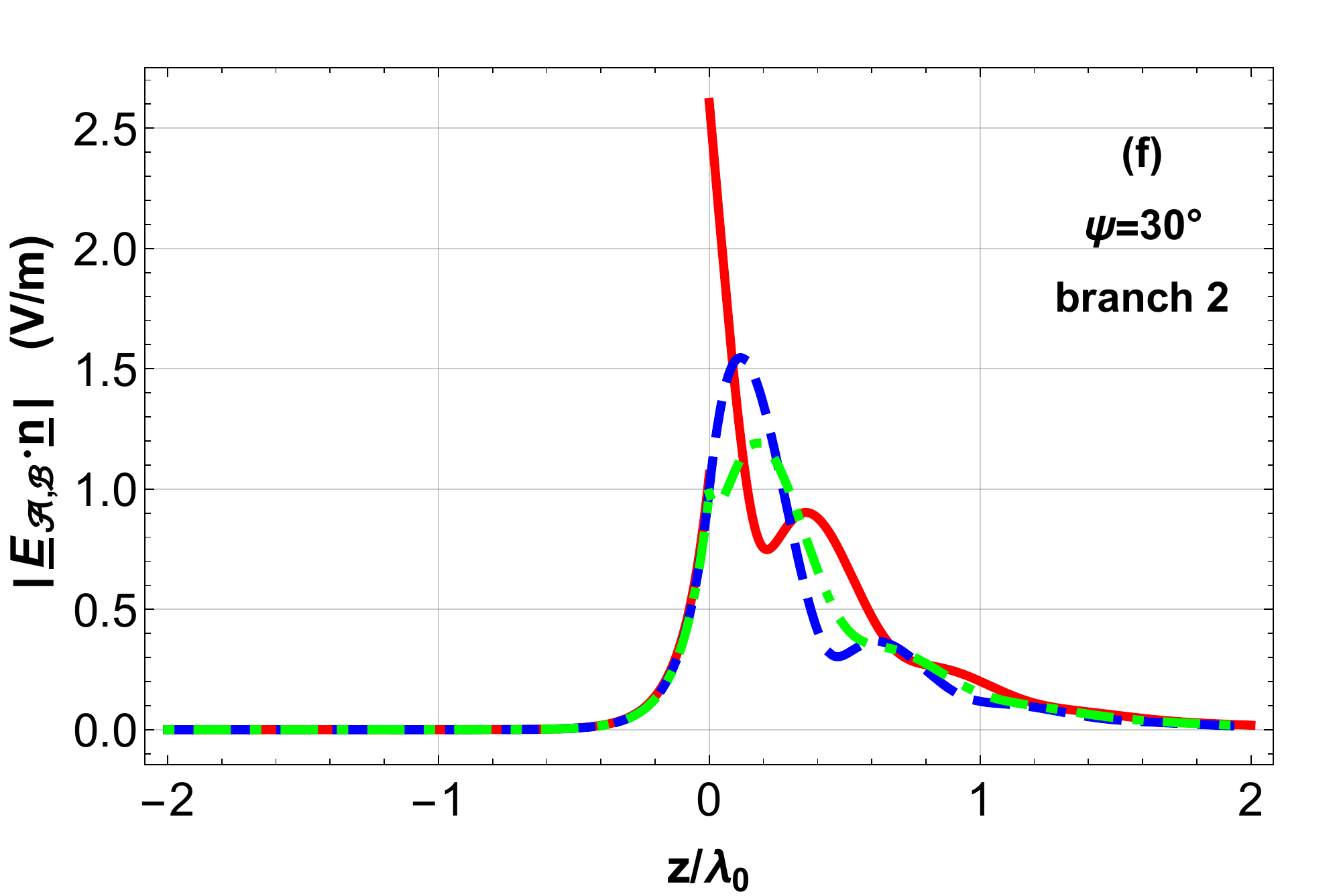}
\end{center}
 \caption{As Fig.~\ref{fig_T260} except that  $T= 290$ K. Here
 $\eps_\mathcal{A}^s =  -2.1565 + 0.2624 i $, $\eps_\mathcal{A}^t =  2.4569 + 0.0556 i $, and
  $\eps_\mathcal{B} =  -5.3130 + 0.5248 i $.
Solution branch 1 (red dashed curves)  exists for $\psi \in \le 0^\circ, 48^\circ \ri$, solution branch 2 (blue broken dashed curves)   exists for $\psi \in \le 0^\circ, 
44^\circ\ri$, and solution branch 3 (green solid curves)  exists for $\psi \in \le 0^\circ, 2^\circ \ri$.
 The quantities  $| \underline{E}_{\, \mathcal{A},\mathcal{B}} (z\hat{\underline{u}}_{\,z}) \. \#n|$ are
 plotted versus $z/\lambdao$ for the branch-2 solution at $\psi = 30^\circ$.
 } \label{fig_T290}
\end{figure}

 By raising the temperature, the regime is reached in which
 $\mbox{Re} \lec \eps^t_\calA \ric > 0  >  \mbox{Re} \lec \eps^s_\calA \ric > \mbox{Re} \lec \eps_\calB \ric  $; i.e., material \red{$\calB$} is a plasmonic material
 while material \red{$\calA$} is a hyperbolic material  (since the product $ \mbox{Re} \lec \eps^s_\calA \ric  \mbox{Re} \lec \eps^t_\calA \ric <0 $). For example, at
  $T = 290$ K with   $f^\calB_{\text{InSb}} = 1$,  the relative-permittivity parameters  are
  $\eps_\mathcal{A}^s =  -2.1565 + 0.2624 i $, $\eps_\mathcal{A}^t =  2.4569 + 0.0556 i $, and
  $\eps_\mathcal{B} =  -5.3130 + 0.5248 i $. 
Plots corresponding to those in Fig.~\ref{fig_T260}  are  provided in Fig.~\ref{fig_T290} for  $T = 290$ K.
In this case there are three solution branches:
 branch 1   exists for $\psi \in \le 0^\circ, 48^\circ \ri$,  branch 2     for $\psi \in \le 0^\circ, 44^\circ\ri$, and
 branch 3    for $\psi \in \le 0^\circ, 2^\circ \ri$. 
The      profiles of    $|\underline{E}_{\, \ell} (z\uz) \. \#n|$,
$\ell \in \lec \mathcal{A}, \mathcal{B} \ric$,
   provided in Fig.~\ref{fig_T290} are for the branch-2 solution at $\psi = 30^\circ$. (The field profiles for the branch-1 solution are provided later in Figs.~\ref{Fields_30deg} and \ref{Fields_5deg}.)
 As in Figs.~\ref{fig_T260} and \ref{fig_T260_E0}, the field  profiles in Fig.~\ref{fig_T290} reveal an
  approximately exponential decay  as the distance from the interface $z=0$ increases, in both the  $+z$ and   $-z$ directions, which confirms that
these solutions do indeed represent surface waves that are localized to the interface $z=0$.

 Most conspicuously, the real part of the  propagation constant $q$ for the branch-1 solution is negative   for $\psi > 11.5^\circ$
 (and positive   for $\psi < 11.5^\circ$). Thus,  for $\psi > 11.5^\circ$  the phase of the branch-1 surface wave propagates in the direction of $- \ux$ whereas the 
 surface wave attenuates in the direction of $+ \ux$.
 Furthermore, since $\mbox{Re} \lec q \ric$  varies uniformly as the angle $\psi$ is varied, as $\psi$ approaches  $ 11.5^\circ$ the phase speed of the branch-1 surface wave becomes  unbounded because  $\mbox{Re} \lec q \ric \approx 0$.
 In contrast, $\mbox{Re} \lec q \ric > 0$ for the branch-2 and branch-3 solutions for all values of $\psi$.
  The magnitudes of $\mbox{Re} \lec q \ric$ are less than unity on all branches, so the phase speeds are greater than the phase speed of a plane wave in free space for all solutions.
 In general, $\mbox{Im} \lec q \ric$,  $\mbox{Re} \lec \alpha_{\calA1} \ric $, $\mbox{Re} \lec \alpha_{\calA2} \ric $, and $\mbox{Re} \lec \alpha_\calB \ric$ 
  in Fig.~\ref{fig_T290} are slightly larger than the corresponding values presented in Fig.~\ref{fig_T260}.
 
\begin{figure}[!htb]
\begin{center}
\includegraphics[width=7.5cm]{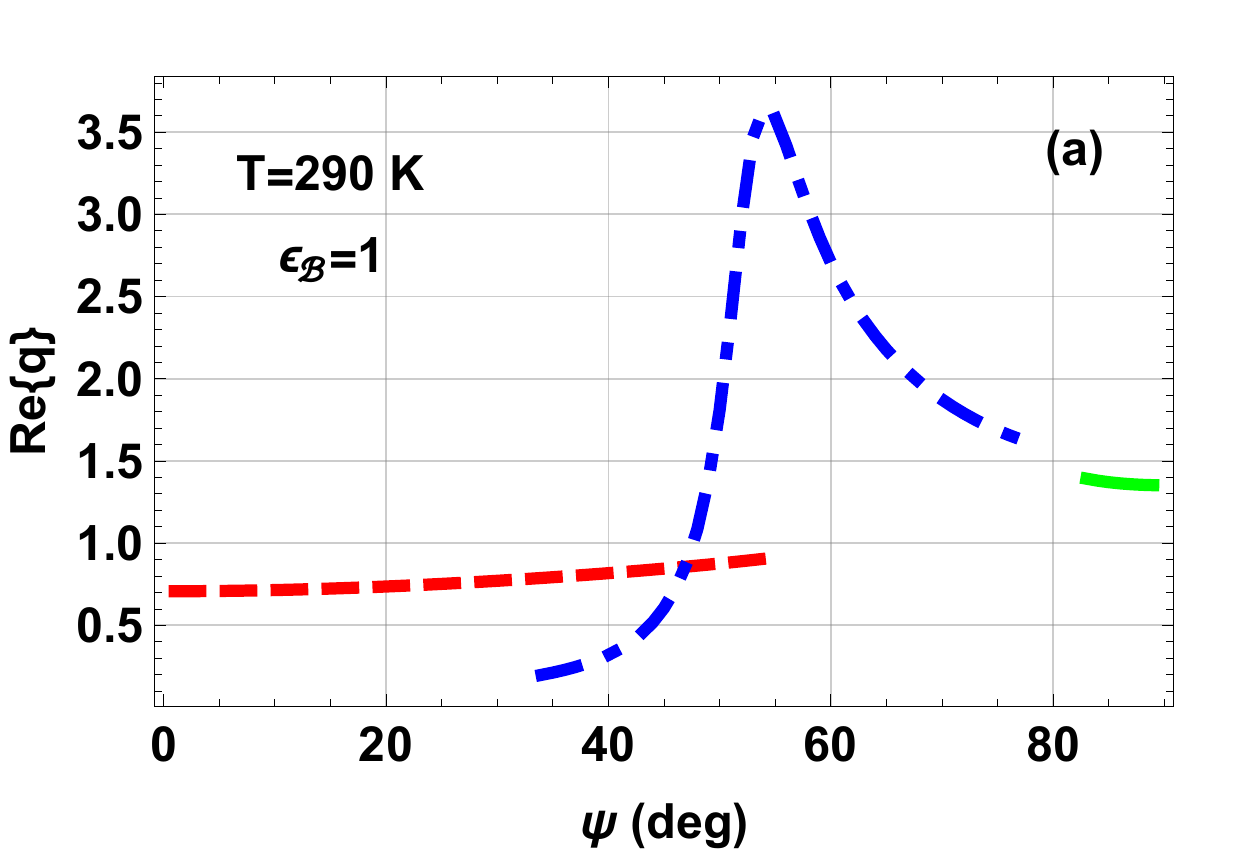} \hfill
\includegraphics[width=7.5cm]{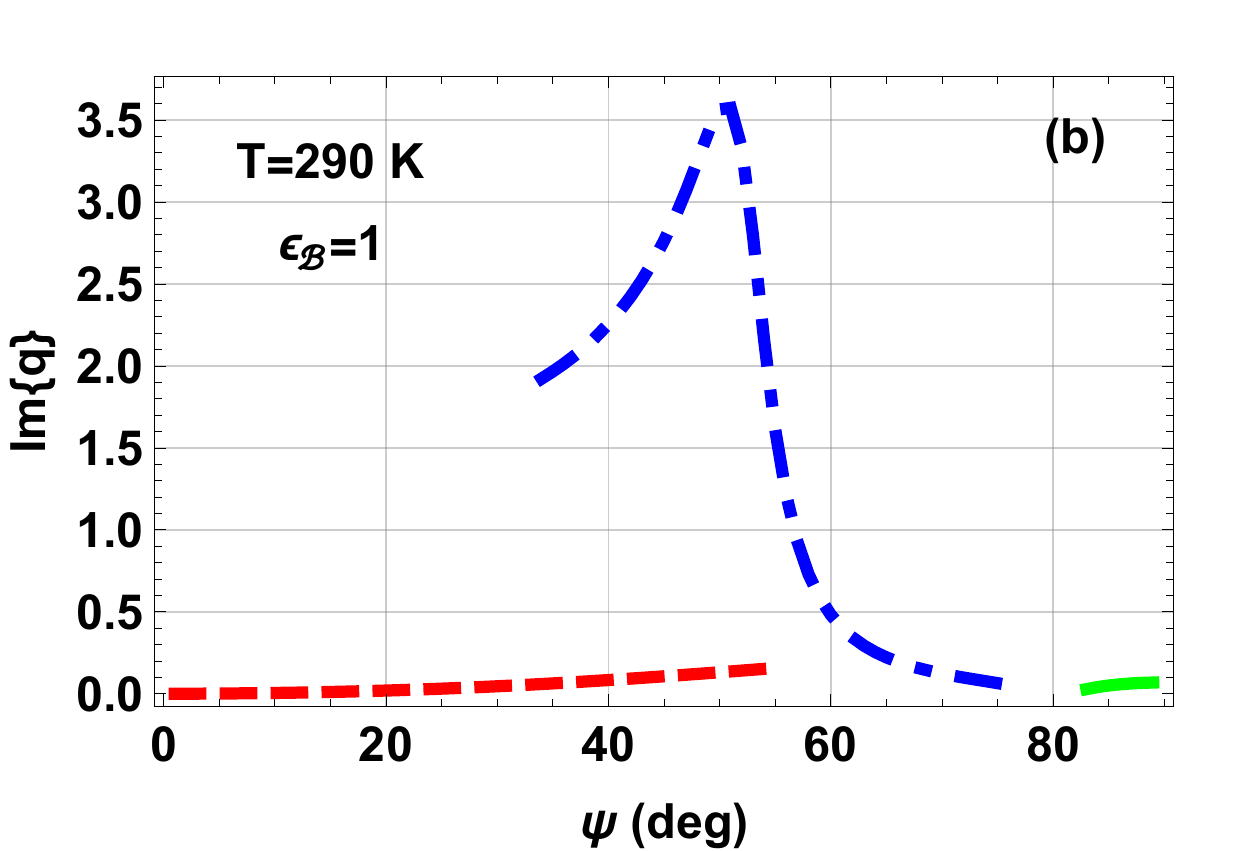} \vspace{4mm} \\
\includegraphics[width=7.5cm]{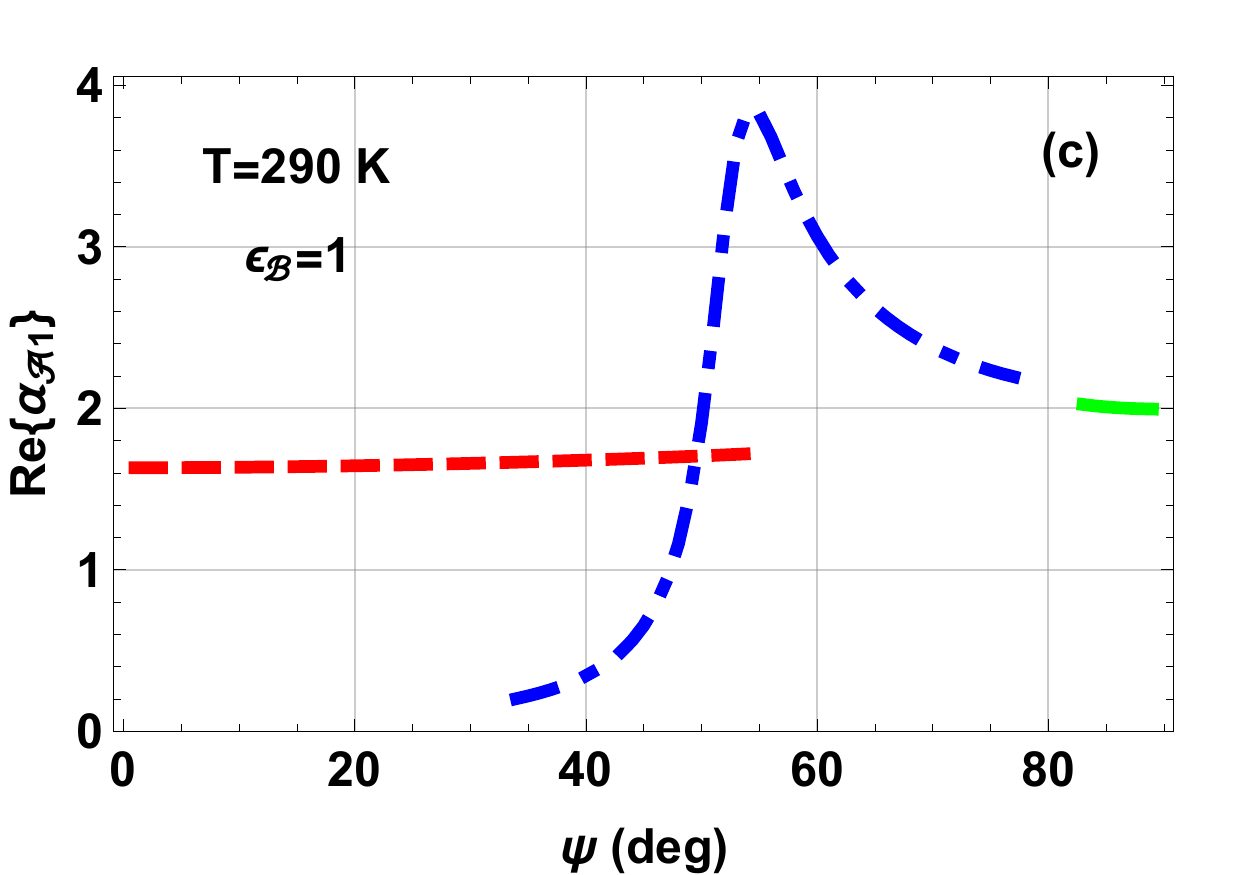} \hfill
\includegraphics[width=7.5cm]{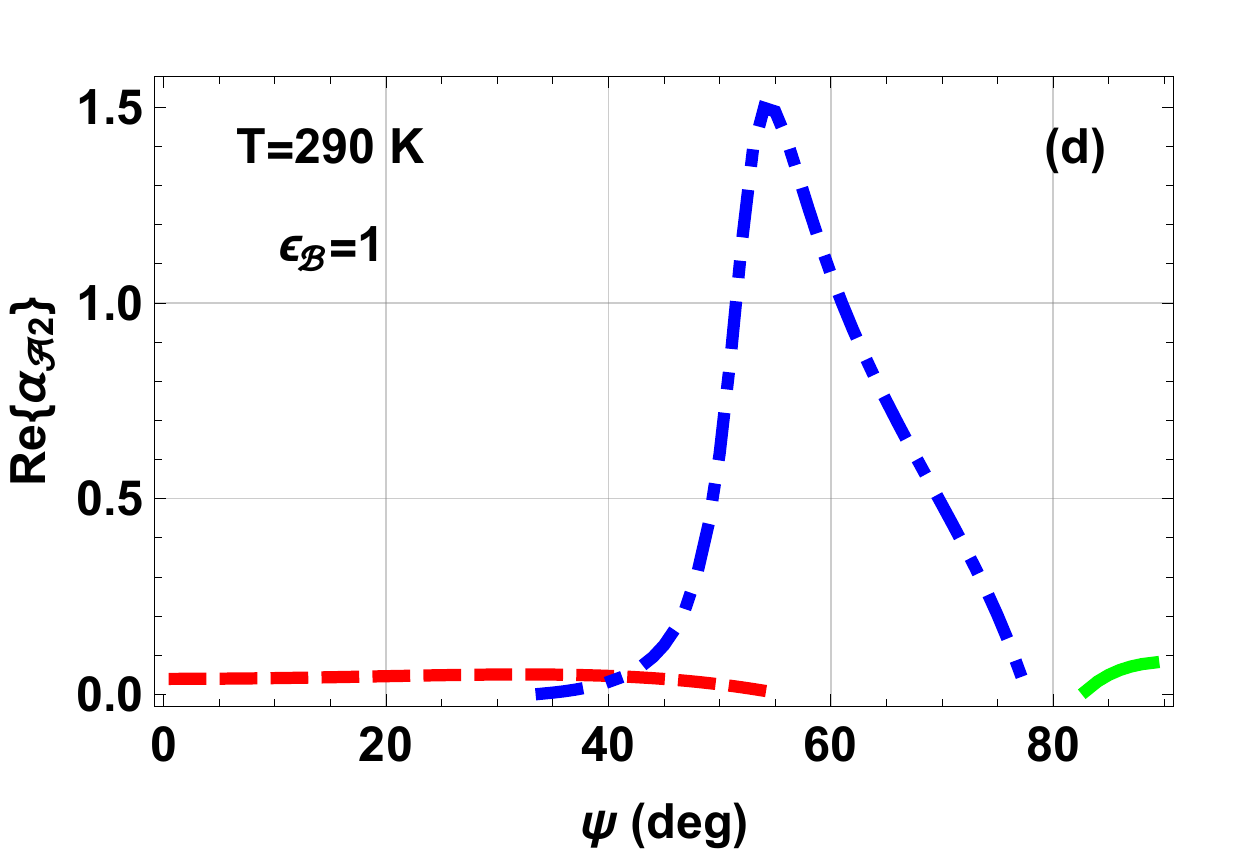} \vspace{4mm} \\
\includegraphics[width=7.5cm]{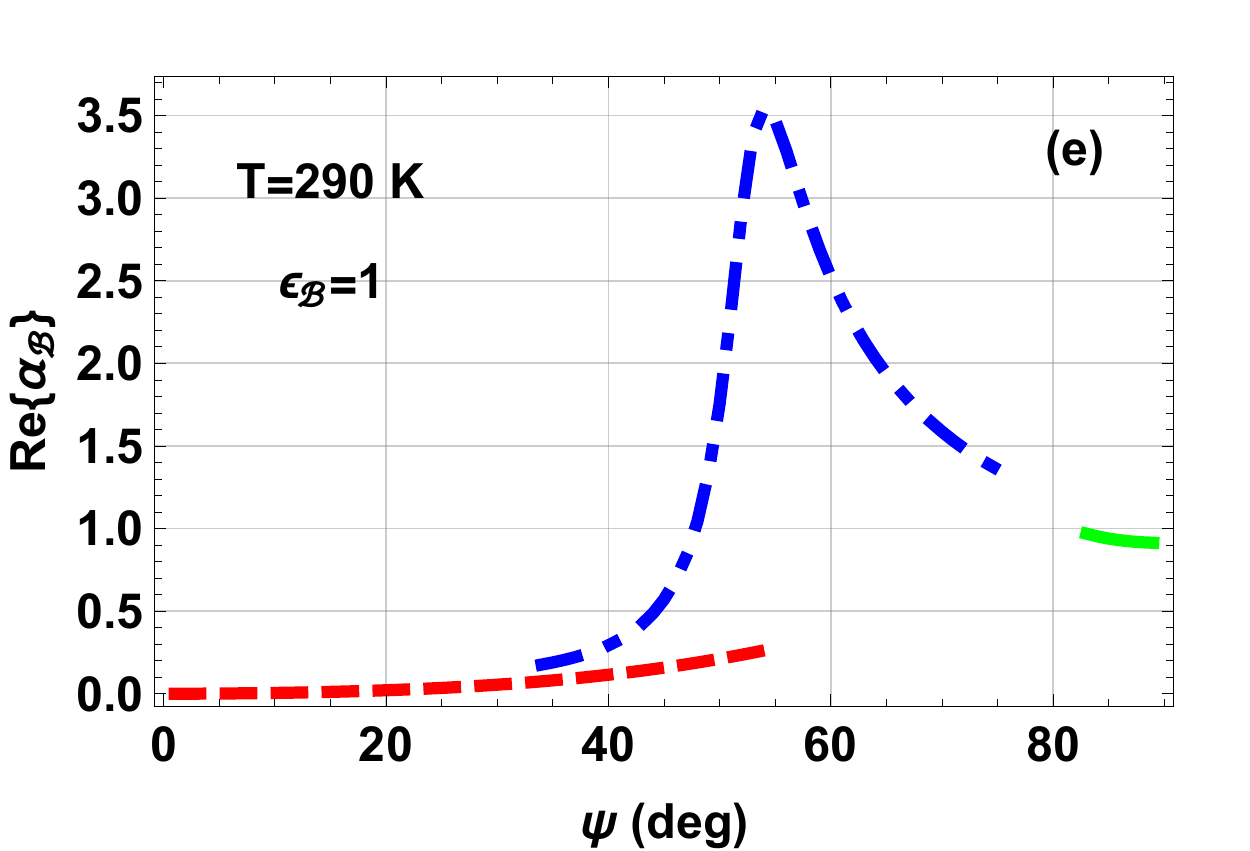}  \hfill
\includegraphics[width=7.5cm]{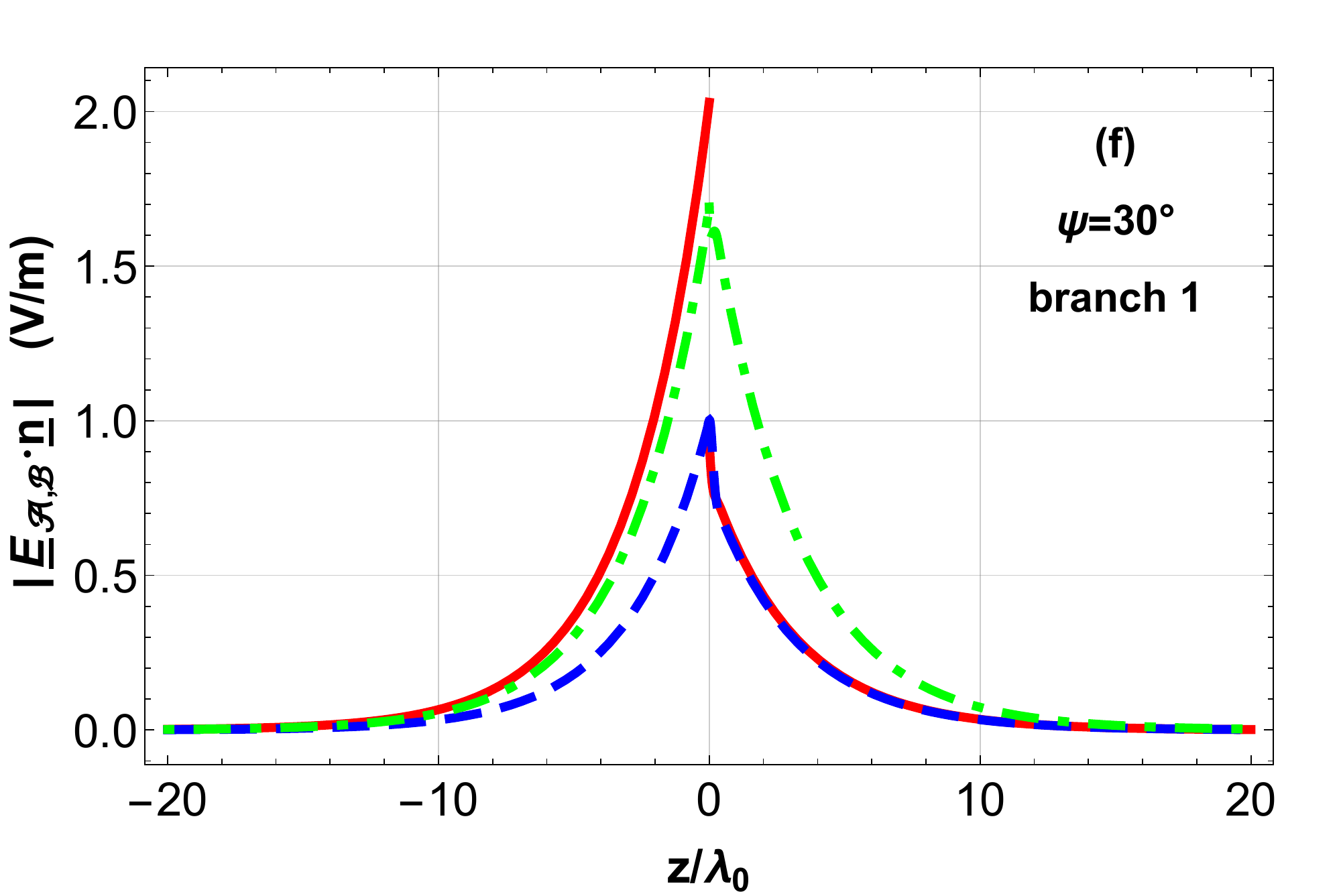}
\end{center}
 \caption{As Fig.~\ref{fig_T290} except that    $\eps_\calB = 1$.
 Solution branch 1 (red dashed curves)  exists for $\psi \in \le 0^\circ, 55^\circ \ri$, solution branch 2 (blue broken dashed curves)   exists for $\psi \in \le 34^\circ, 
77^\circ\ri$, and solution branch 3 (green solid curves)  exists for $\psi \in \le 83^\circ, 90^\circ \ri$.
 The quantities  $| \underline{E}_{\, \mathcal{A},\mathcal{B}} (z\hat{\underline{u}}_{\,z}) \. \#n|$ are
 plotted versus $z/\lambdao$ for the branch-1 solution at $\psi = 30^\circ$.
 } \label{fig_T290_E0}
\end{figure}

 The results in  Fig.~\ref{fig_T290} can be compared with the corresponding results that arise when material $\calB$ is simply air, i.e., $\eps_\calB= 1$,
 which are presented in   Fig.~\ref{fig_T290_E0}. As in  Fig.~\ref{fig_T290}, there are three  solution branches for  Fig.~\ref{fig_T290_E0}: 
  branch 1   exists for $\psi \in \le 0^\circ, 55^\circ \ri$,  branch 2    for $\psi \in \le 34^\circ, 77^\circ\ri$, and
 branch 3   for $\psi \in \le 83^\circ, 90^\circ \ri$. For all three  solution branches, the phase speeds are small relative to the phase speed of a plane wave in free space. Unlike Fig.~\ref{fig_T290}, $\mbox{Re} \lec q \ric < 0$ does not exist in Fig.~\ref{fig_T290_E0}.
The      profiles of    $|\underline{E}_{\, \ell} (z\uz) \. \#n|$,
$\ell \in \lec \mathcal{A}, \mathcal{B} \ric$,
   provided in Fig.~\ref{fig_T290_E0} are for the branch-1 solution at $\psi = 30^\circ$. 
As compared with the
  $T = 260$ K case of  Fig.~\ref{fig_T260_E0}, the surface waves represented in 
  Fig.~\ref{fig_T290_E0}
   decay much more rapidly  in directions both parallel and normal to the interface $z=0$.

  \subsection{Negative phase velocity}
  
\begin{figure}[!htb]
\begin{center}\includegraphics[width=7.6cm]{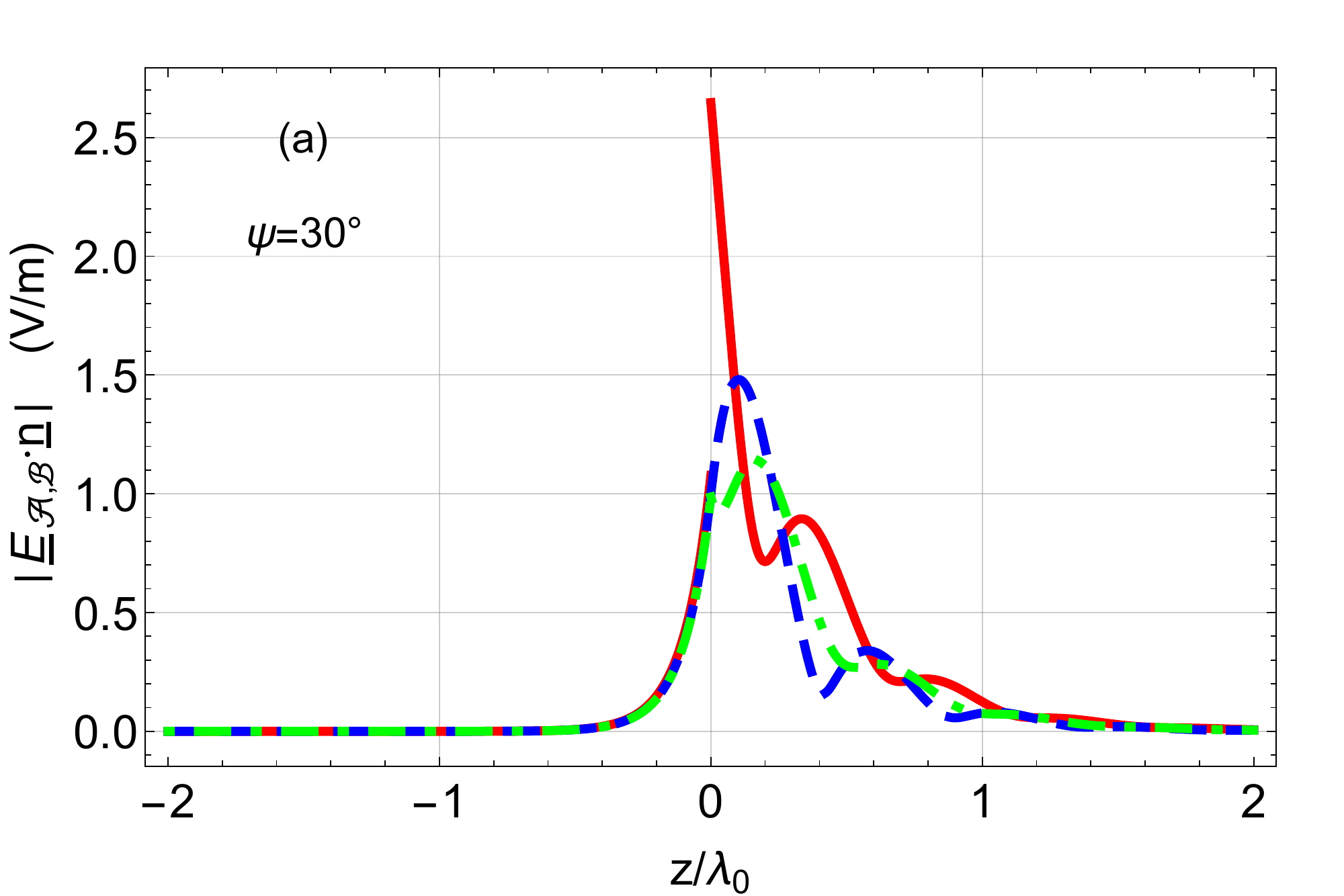} \hfill
\includegraphics[width=7.6cm]{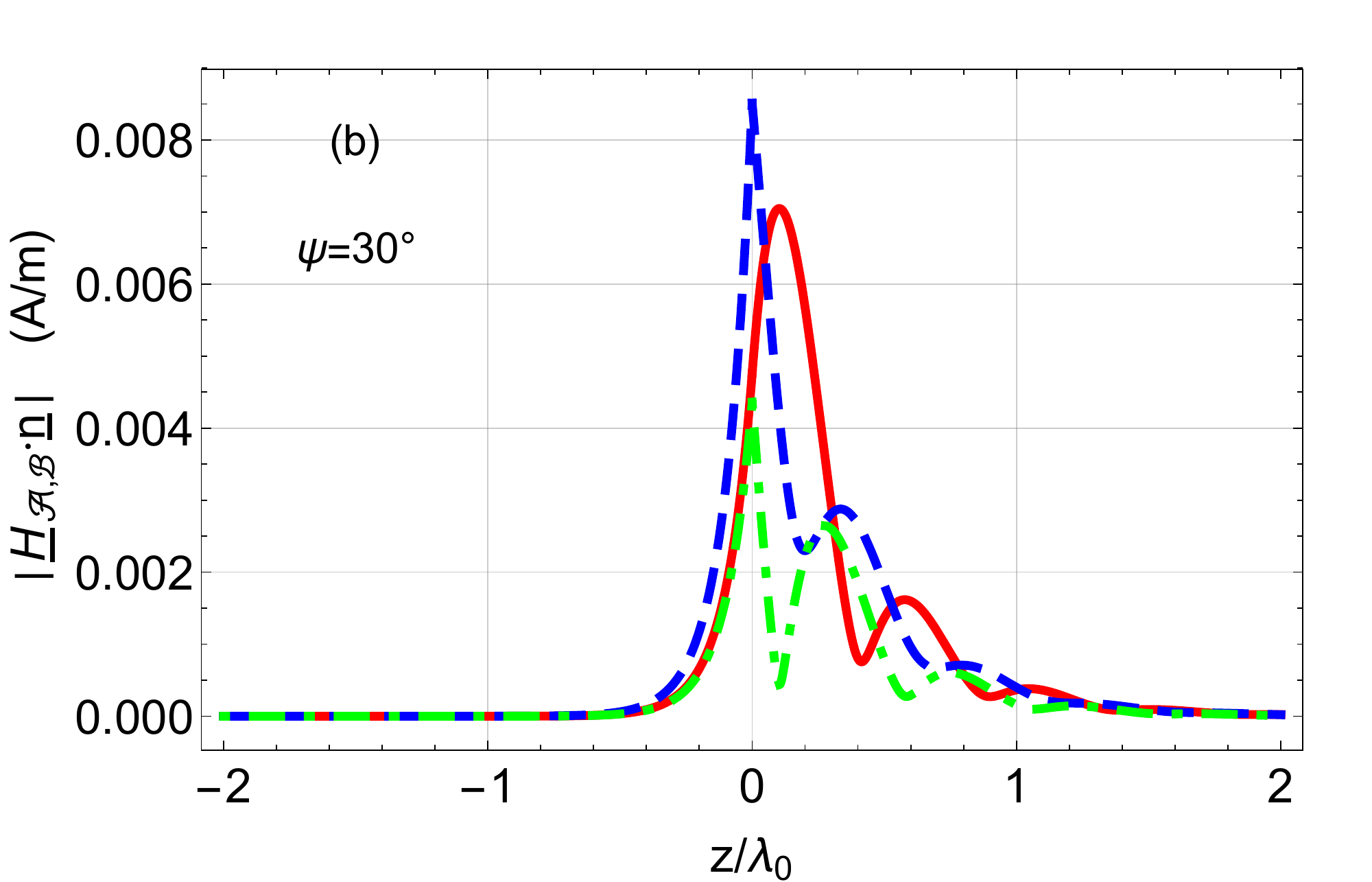} \vspace{4mm}  
\includegraphics[width=7.6cm]{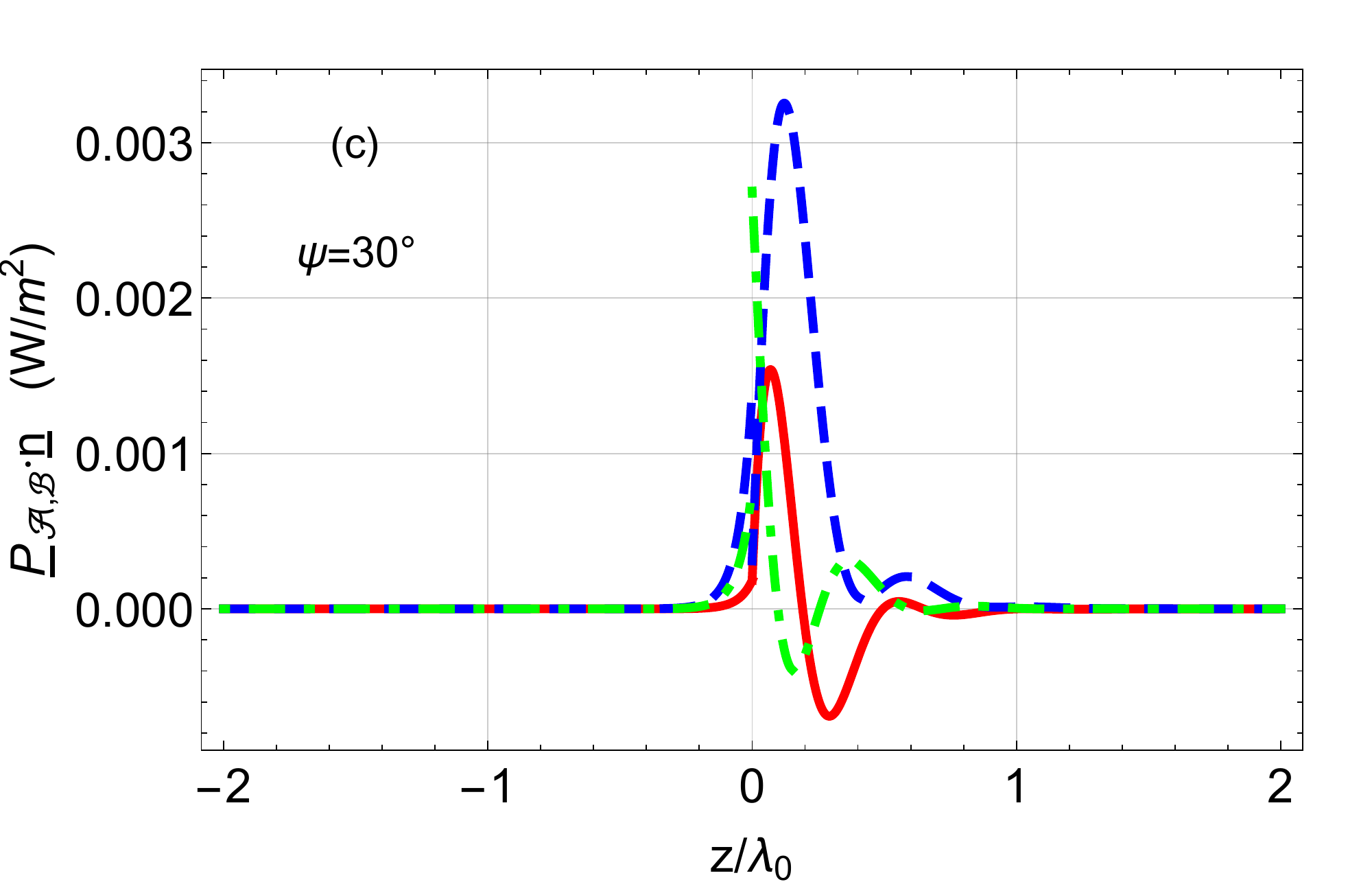}
\hfill
\includegraphics[width=7.6cm]{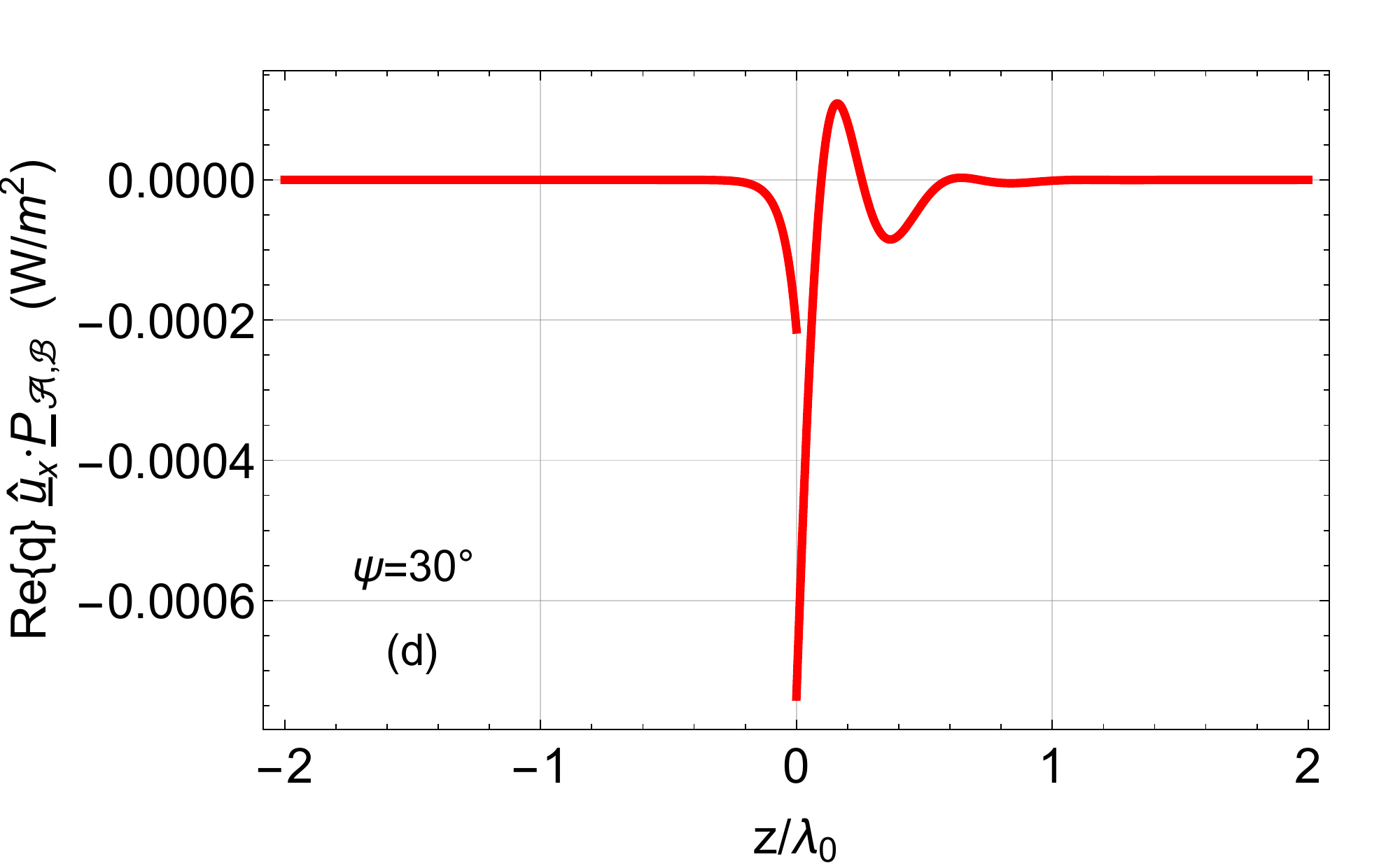}
\end{center}
 \caption{ The quantities (a) $| \underline{E}_{\, \mathcal{A},\mathcal{B}} (z\hat{\underline{u}}_{\,z}) \. \#n|$, 
 (b)  $|\underline{H}_{\, \mathcal{A},\mathcal{B}}  (z\hat{\underline{u}}_{\,z}) \. \#n|$, (c) $\underline{P}_{\, \mathcal{A},\mathcal{B}}  (z\hat{\underline{u}}_{\,z}) \. \#n$, 
 and (d) $\mbox{Re} \lec q \ric \ux \. \underline{P}_{\, \mathcal{A},\mathcal{B}}  (z\hat{\underline{u}}_{\,z}) $ 
 plotted versus $z/\lambdao$,
 for the parameter values of Fig.~\ref{fig_T290}, branch 1 solution, 
 with $\psi = 30^\circ$ and $ \#{\mathcal E}_{\,\calB} \. \uy = 1$ V m${}^{-1}$.
 Key:   $\#n = \ux$ broken dashed green curves; $\#n = \uy$ dashed blue curves; $\#n = \uz$ solid red curves.
 } \label{Fields_30deg}
\end{figure}

  Let us return to the issue of  $\mbox{Re} \lec q \ric < 0$   which arises in Fig.~\ref{fig_T290}
  for branch 1 when  $\psi > 11.5^\circ$.
For further exploration,
   profiles of    $|\underline{E}_{\, \ell} (z\uz) \. \#n|$
and $|\underline{H}_{\, \ell} (z\uz) \. \#n|$,
$\ell \in \lec \mathcal{A}, \mathcal{B} \ric$,
 versus $z / \lambdao$ for $\#n \in \lec \ux, \uy, \uz \ric$ are plotted
 in Fig.~\ref{Fields_30deg},  using the same parameter values as for the branch-1 solution in  Fig.~\ref{fig_T290}
with $\psi = 30^\circ$ and $\#{\mathcal E}_{\,\calB} \. \uy = 1$ V m${}^{-1}$.  In addition, plots are provided of  the corresponding profiles of the Cartesian components 
$\underline{P}_{\, \ell} (z\uz) \. \#n$, $\ell \in \lec \mathcal{A}, \mathcal{B} \ric$ and $\#n \in \lec \ux, \uy, \uz \ric$, of the time-averaged Poynting vector
\begin{equation}
\underline{P}_{\, \ell} (\#r) = \frac{1}{2} \mbox{Re} \les \, \underline{E}_{\, \ell} (\#r)  \times 
\underline{H}^*_{\, \ell} (\#r) \, \ris, \qquad \ell \in \lec  \mathcal{A}, \mathcal{B}  \ric\,,
\end{equation}
where the asterisk denotes the complex conjugate. The dot product $\mbox{Re} \lec q \ric \ux \. \underline{P}_{\, \mathcal{A},\mathcal{B}}  (z\hat{\underline{u}}_{\,z}) $ is plotted  versus $z / \lambdao$ too. 

Most significantly, $\mbox{Re} \lec q \ric \ux \. \underline{P}_{\, \mathcal{A},\mathcal{B}}  (z\hat{\underline{u}}_{\,z}) < 0$
for much of the range of $z$  considered.  Satisfaction of the inequality $\mbox{Re} \lec q \ric \ux \. \underline{P}_{\, \mathcal{A},\mathcal{B}}  (z\hat{\underline{u}}_{\,z}) < 0$  signifies  NPV.
The  surface wave is localized to the   interface $z=0$, albeit 
there is significant spreading of the fields into
 both the half-spaces $z>0$ and $z<0$, but the degree of localization is greater in the half-space $z<0$ than in the half-space $z>0$.

\begin{figure}[!htb]
\begin{center}\includegraphics[width=7.6cm]{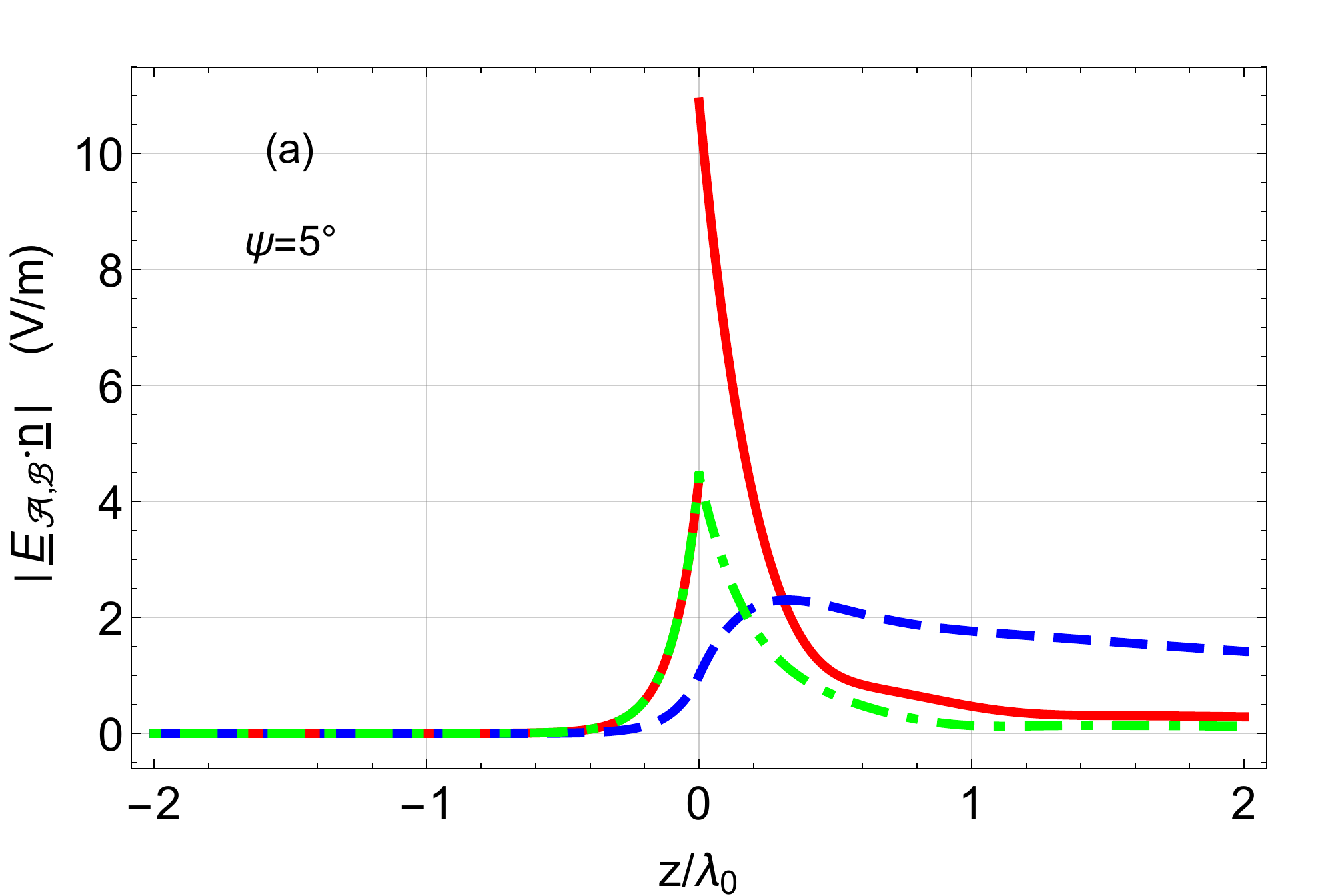} \hfill
\includegraphics[width=7.6cm]{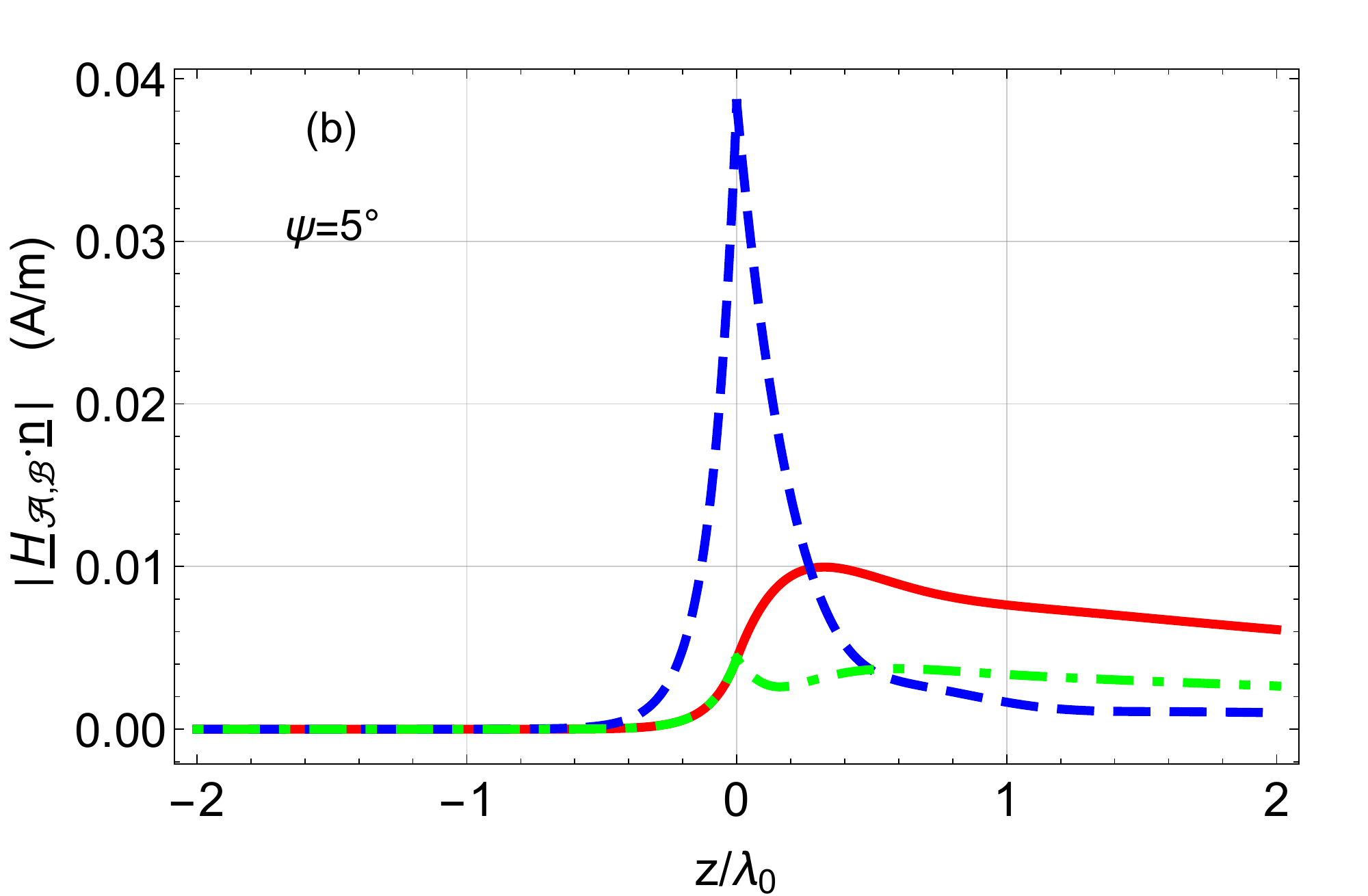} \vspace{4mm}  
\includegraphics[width=7.6cm]{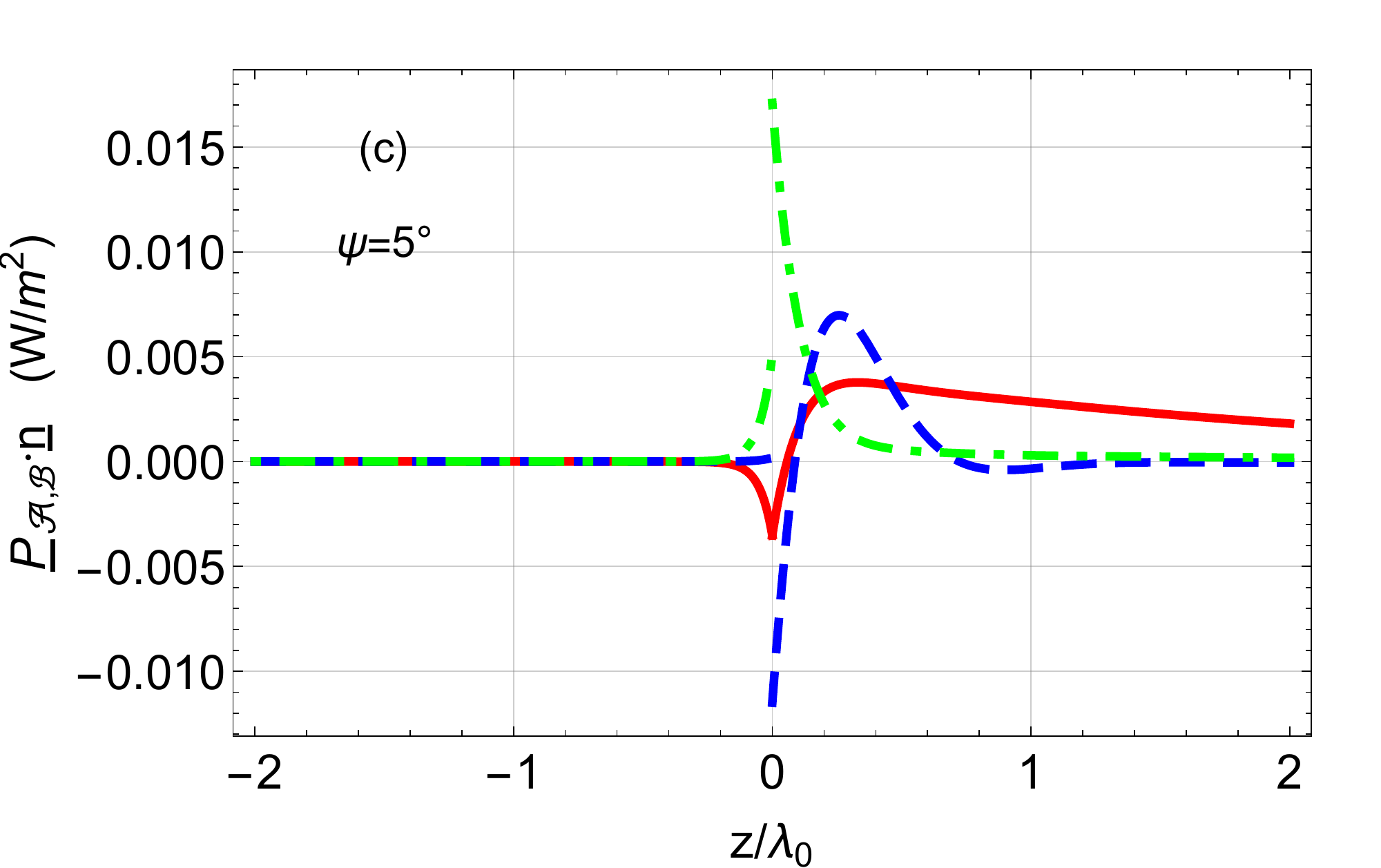}
\hfill
\includegraphics[width=7.6cm]{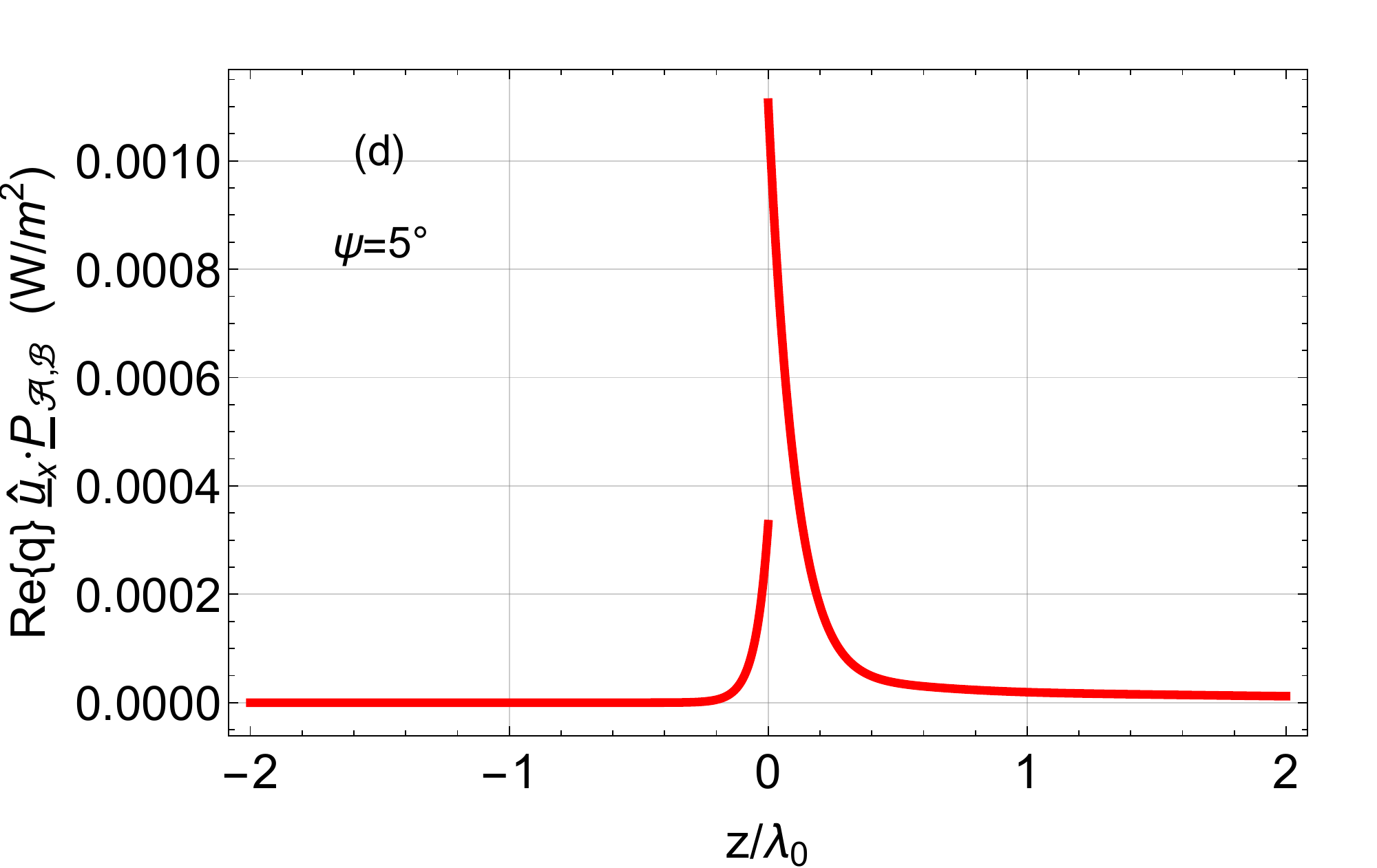}
\end{center}
 \caption{As Fig.~\ref{Fields_30deg} but for $\psi = 5^\circ$.
 } \label{Fields_5deg}
\end{figure}

For comparison, the analogous plots for $\psi = 5^\circ$ are provided in Fig.~\ref{Fields_5deg}. In this case,  $\mbox{Re} \lec q \ric > 0$
  and  $\mbox{Re} \lec q \ric \ux \. \underline{P}_{\, \mathcal{A},\mathcal{B}}  (z\hat{\underline{u}}_{\,z}) > 0$
for the entire $z$ range considered. Thus, there is no NPV when $\psi = 5^\circ$.   The degree of localization at the   interface $z=0$ is similar for
  Figs.~\ref{Fields_30deg} and \ref{Fields_5deg}, for both $z<0$ and $z>0$.

\begin{figure}[!htb]
\begin{center}\includegraphics[width=14cm]{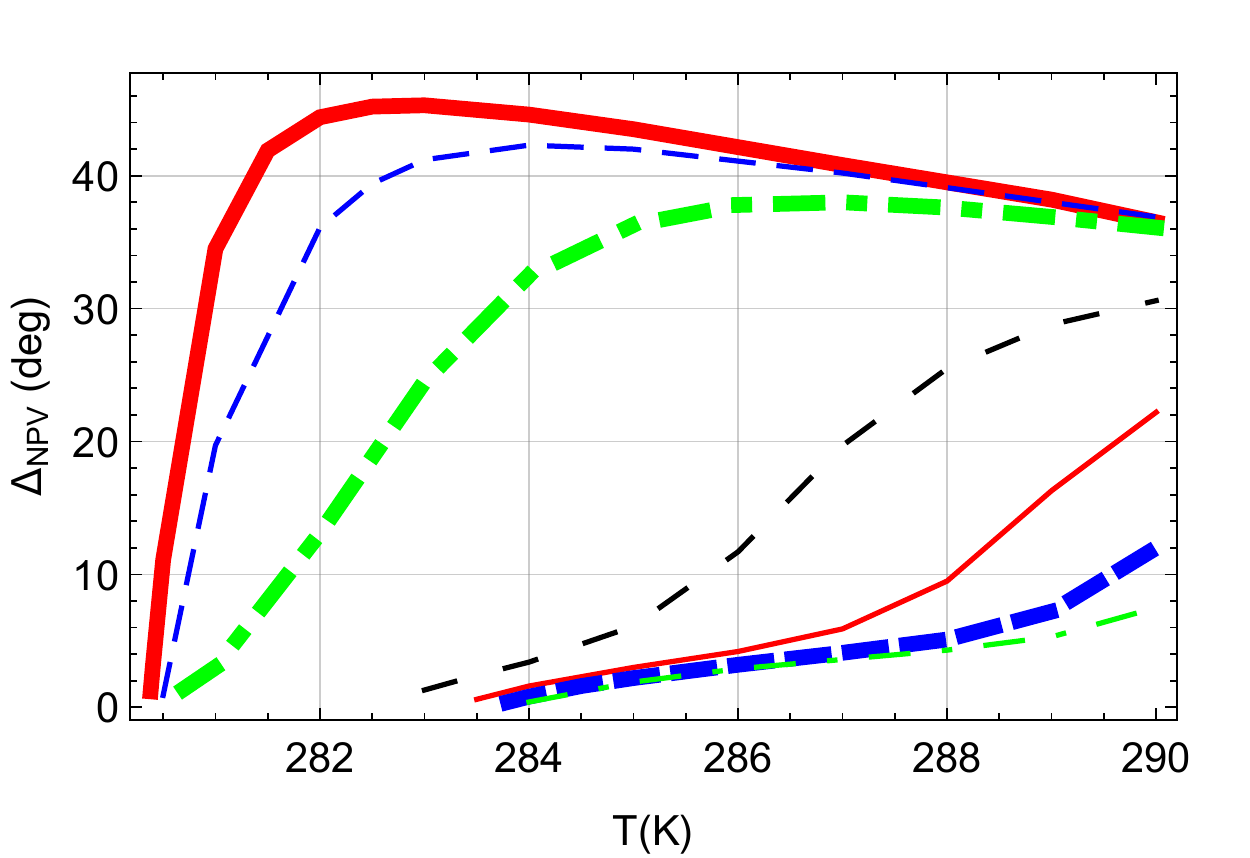} 
\end{center}
 \caption{$\Delta \psi_{\text{NPV}}$ plotted against   plotted against  $T \in \le 280, 290 \ri $ K for  $f^\calB_{\text{InSb}} = 1$ (solid thick red curve), 
 0.8 (dashed thin blue curve), 0.7 (broken dashed thick green curve), 0.62 (dashed with long spaces, thin black  curve), 0.6 (solid thin red curve), 0.59 (dashed thick blue curve) and 
 0.585 (broken dashed thin green curve). } \label{AED_T}
\end{figure}

 Let us denote the extent of the continuous range of angular directions  over which NPV is supported by the branch-1 solution in Fig.~\ref{fig_T290} by $\Delta \psi_{\text{NPV}}$. For the parameter values used  in Fig.~\ref{fig_T290},  $\Delta \psi_{\text{NPV}} = 36.5^\circ$.
  In Fig.~\ref{AED_T},
  $\Delta \psi_{\text{NPV}}$ plotted against   plotted against  $T \in \le 260, 290 \ri $ K for  $f^\calB_{\text{InSb}} \in \lec
  0.585, 0.59, 0.6, 0.62, 0.7,  0.8,  1 \ric$. At a fixed temperature,   $\Delta \psi_{\text{NPV}}$ increases uniformly as $f^\calB_{\text{InSb}} $ increases.
  At a fixed volume fraction   $f^\calB_{\text{InSb}}$,  $\Delta \psi_{\text{NPV}}$ varies substantially as $T$ increases, depending upon the value of 
  $f^\calB_{\text{InSb}}$. The lowest temperature at which NPV solutions are supported increases uniformly as  $f^\calB_{\text{InSb}}$ decreases.

\section{Closing remarks}

Hyperbolic materials are associated with a host of exotic electromagnetic phenomenons \c{Smolyaninov}. Our numerical studies have
 revealed that such materials, for certain constitutive-parameter  ranges, can support the propagation of surface waves with NPV.
 Furthermore, hyperbolic materials can support a multiplicity of surface waves in a given propagation direction. One of these waves may exhibit NPV while the others exhibit positive phase velocity.
In our numerical investigations, NPV was only found to be supported, for certain constitutive-parameter  ranges, when 
partnering material $\calA$ was hyperbolic and
partnering material $\calB$ was plasmonic. That is, the conditions 
\begin{equation}
\left.
\begin{array}{l}
\mbox{Re} \lec \eps_\mathcal{A}^s  \ric \mbox{Re} \lec \eps_\mathcal{A}^t  \ric < 0  \vspace{6pt} \\
\mbox{Re} \lec \eps_\mathcal{B} \ric < 0
\end{array}
\right\}
\end{equation}
are necessary conditions for surface-wave NPV, but not sufficient conditions.

\vspace{10mm}

\noindent {\bf Acknowledgments.}
TGM acknowledges the support of EPSRC grant EP/S00033X/1.
AL thanks the Charles Godfrey Binder Endowment at the Pennsylvania State University for ongoing support of his research, with partial funding from US NSF grant number DMS-1619901.

\newpage

\newpage

\newpage

\newpage

\newpage

\newpage

\newpage

\newpage

\newpage

\end{document}